\documentclass{article}
\usepackage[a4paper,top=3cm,bottom=2cm,left=2.5cm,right=2.5cm,marginparwidth=1.75cm]{geometry}
\usepackage{array}
\usepackage[utf8]{inputenc}
\usepackage[round]{natbib}
\defcitealias{WMO-OSCAR}{WMO OSCAR, 2023}
\defcitealias{SAC_report}{ECMWF SAC, 2022}
\defcitealias{EUCOS_report}{EUMETNET report, 2018}
\defcitealias{TIGGE-LAM}{TIGGE-LAM dataset, 2023}

\usepackage{amssymb}
\usepackage{amsmath}
\usepackage{url}
\usepackage{diagbox}
\usepackage{authblk}
\usepackage[dvipsnames]{xcolor}
\usepackage{mathabx}

\newcommand\expt[1]{\ensuremath{\mathbb{E}\left[{#1}\right]}}
\newcommand\m[1]{\ensuremath{\mathbf{#1}}}
\newcommand\e[1]{\ensuremath{\mathrm{#1}}}

\newcommand\wh[1]{\ensuremath{\widehat{#1}}}

\newcommand\whm[1]{\ensuremath{\widehat{\mathbf{#1}}}}
\newcommand{\reals}{\ensuremath{\mathbb{R}}}
\newcommand{\err}{\ensuremath{\boldsymbol{\varepsilon}}}

\newcommand{\y}{\Large \textcolor{ForestGreen}{\checkmark}}
\newcommand{\n}{\textcolor{red}{\bigtimes}}
\newcommand{\s}{\textcolor{YellowOrange}{\bigcirc}}

\providecommand{\keywords}[1]
{
  \small	
  \textbf{\textit{Keywords---}} #1
}

\title{On methods for assessment of the influence and impact of observations in convection-permitting numerical weather prediction}

\author[1,2]{Guannan Hu \thanks{guannan.hu@reading.ac.uk}}
\author[1,2,3]{Sarah L. Dance}
\author[1,2]{Alison Fowler}
\author[4]{David Simonin}
\author[4]{Joanne Waller}
\author[5]{Thomas Auligne}
\author[6]{Sean Healy}
\author[6,7]{Daisuke Hotta}
\author[8,9]{Ulrich L\"{o}hnert}
\author[10,11,12]{Takemasa Miyoshi}
\author[13,14]{Nikki C. Priv\'{e}}
\author[15]{Olaf Stiller}
\author[16]{Xuguang Wang}
\author[17]{Martin Weissmann}

\affil[1]{Department of Meteorology, University of Reading, United Kingdom}
\affil[2]{National Centre for Earth Observation, University of Reading, United Kingdom}
\affil[3]{Department of Mathematics and Statistics, University of Reading, United Kingdom}
\affil[4]{Met Office, Reading, UK}
\affil[5]{University Corporation for Atmospheric Research, Joint Center for Satellite Data Assimilation, Boulder, Colorado, USA}
\affil[6]{European Centre for Medium-Range Weather Forecasts, Reading, United Kingdom}
\affil[7]{Meteorological Research Institute, Japan Meteorological Agency, Tsukuba, Japan}
\affil[8]{Institute for Geophysics and Meteorology, University of Cologne, Cologne, Germany}
\affil[9]{Hans-Ertel-Centre for Weather Research, Climate Monitoring and Diagnostics, Cologne/Bonn, Germany}
\affil[10]{RIKEN Center for Computational Science, Kobe, 6500047, Japan}
\affil[11]{RIKEN Cluster for Pioneering Research, Kobe, 6500047, Japan}
\affil[12]{RIKEN Interdisciplinary Theoretical and Mathematical Sciences Program (iTHEMS), Kobe, 6500047, Japan}
\affil[13]{Morgan State University/GESTAR II, Baltimore, Maryland}
\affil[14]{Global Modeling and Assimilation Office, NASA, Greenbelt, Maryland}
\affil[15]{Data Assimilation Section, Deutscher Wetterdienst (DWD), Frankfurt am Main, 63067, Germany}
\affil[16]{School of Meteorology, University of Oklahoma, Norman, Oklahoma}
\affil[17]{Institut für Meteorologie und Geophysik, Universität Wien, Vienna, Austria}

\date{}

\begin{document}

\maketitle

\begin{abstract}
    In numerical weather prediction (NWP), a large number of observations are used to create initial conditions for weather forecasting through a process known as data assimilation. An assessment of the value of these observations for NWP can guide us in the design of future observation networks, help us to identify problems with the assimilation system, and allow us to assess changes to the assimilation system. However, the assessment can be challenging in convection-permitting NWP. First, the strong nonlinearity in the forecast model limits the methods available for the assessment. Second, convection-permitting NWP typically uses a limited area model and provides short forecasts, giving problems with verification and our ability to gather sufficient statistics.  Third, convection-permitting NWP often makes use of novel observations, which can be difficult to simulate in an observing system simulation experiment (OSSE). We compare methods that can be used to assess the value of observations in convection-permitting NWP and discuss operational considerations when using these methods. We focus on their applicability to ensemble forecasting systems, as these systems are becoming increasingly dominant for convection-permitting NWP. We also identify several future research directions: comparison of forecast validation using analyses and observations, the effect of ensemble size on assessing the value of observations, flow-dependent covariance localization, and generation and validation of the nature run in an OSSE.
\end{abstract}
 

\section{Introduction}\label{sec:introduction}

Convection-permitting numerical weather prediction (NWP) is essential for predicting high-impact weather events, such as heavy precipitation, storms, floods, wind gusts and fog \citep[e.g.,][]{Dance2019,Clark2016}. Convection-permitting models typically have a small horizontal grid-length (around 1-4 km), allowing convection to be explicitly modelled rather than parameterized \citep{Hu2022}, although research models do run at a smaller grid lengths \citep{Waller2021,miyamoto2013deep}. Compared to convection-permitting models, convection-resolving models should have an even smaller grid length \citep{Prein2015}. Crucial to the realism of NWP at any scale is the routine assimilation of data. The data assimilation (DA) process combines observations with a short-range model forecast (called the background) to provide the initial conditions (called the analysis) for the NWP model to produce weather forecasts \citep[e.g.,][]{MilanEtAl2020,Schraff2016}.

Despite the assimilation of observations being essential to forecast skill, it is known that not every observation assimilated reduces the forecast error. This is due to the nature of the random errors present in the observations and the suboptimality in the assimilation system (see \citet{Lorenc2014} for some suggestions for the causes). In addition, the same type of observations may be ranked differently in terms of their impact on global forecasts and regional convection-permitting forecasts. For global NWP, the observation types with the largest overall positive impact are satellite microwave and infrared sounders and radiosondes, while for convection-permitting NWP they are currently radar and aircraft observations \citep{WMOreport}, although novel and unused observations (e.g. ground-based and satellite microwave links or uncrewed aerial systems [UAS]) bear a high potential for further impact (see section \ref{sec:new observation types}). Therefore, the value of different observations (e.g., different types, or from different sensors or channels), as well as the value of different deployments of observations (location, spatial density and temporal frequency) for convection-permitting NWP, need to be properly assessed.

Generally speaking, there are three distinct needs for assessing the value of observations in NWP:

\begin{itemize}
    \item First, there is a need to provide guidance for the maintenance of current observation networks and the design of future observation networks, e.g., the design of new observing systems and new networks of currently available observations.
    \item Second, there is a need to identify problems with the assimilation system, e.g., incorrectly specified background and observation error statistics and inaccurate observation operators.
    \item Third, there is a need to evaluate changes to the assimilation system, e.g., tuning background and observation error variances and optimizing localization functions.
\end{itemize}
These types of scientific assessments are typically used to help guide decisions about deployment of limited financial and human resources, such as development of a new satellite mission or staffing for DA system development.  

Since convection-permitting DA differs in many aspects from global DA \citep[e.g.,][]{Hu2022,Dance2019,Gustafsson2018,Bauer2011,Dance2004}, assessing the value of observations in convection-permitting NWP presents unique challenges. We evaluate current state-of-the-art methods (including methodologies and metrics) for assessing the value of observations in convection-permitting NWP. We provide guidance and recommendations for their future use and development. Throughout this paper, we use the term observation \textit{impact} to represent the value of observations to the \textit{forecast} and observation \textit{influence} to represent the contribution of observations to the \textit{analysis}.

The outline of this paper is as follows. In section \ref{sec:Challenges in C-P NWP}, we describe the challenges in the assessment of the value of observations in convection-permitting NWP. In section \ref{sec:available methods}, we introduce currently available methods and discuss the links between them. In section \ref{sec:pros and cons}, we compare the advantages and disadvantages of different methods in terms of their ability to address the challenges described in section \ref{sec:Challenges in C-P NWP}. In section \ref{sec:operational considerations}, we consider the practical aspects of using these methods and make recommendations for the future. In section \ref{sec:summary}, we conclude by summarising the open questions for each method and some general considerations in assessing the value of observations in convection-permitting NWP.

\section{Challenges for assessing the value of observations in convection-permitting NWP}\label{sec:Challenges in C-P NWP}

Numerous studies have been carried out to assess the value of observations in global NWP \citep[e.g.,][]{Prive2021,Cardinali2018,Lorenc2014,Gelaro2010}. However, less attention has been paid to assessing the value of observations in convection-permitting NWP. In this section, we discuss how three main differences between convection-permitting and global NWP pose challenges in the assessment of the value of observations in convection-permitting NWP.

\subsection{Stronger nonlinearities}\label{sec:nonlinearity}

The atmospheric processes modelled in convection-permitting NWP are generally more nonlinear than in global models. \citet{Hohenegger2007} have shown that the tangent-linear approximation breaks down at 1.5 h forecast lead-time when using a regional model with a horizontal resolution of 2.2 km and at 54 h when using a global model with a horizontal spectral resolution of 80 km. The stronger nonlinearity of error growth imposes three challenges in assessing the value of observations. First, it makes methods that rely strongly on the assumption of linear error growth no longer suitable unless there is a skilled perturbation model available (see further discussion of methods in section \ref{sec:available methods}). Second, it may result in greater variability in estimates of the value of observations on an individual analysis or forecast, which increases the difficulty of achieving a statistically significant estimate. We need to distinguish whether changes in the estimates are caused by modifications to the observing system or by the chaotic nature of any weather forecast, which requires sufficient spatial and temporal averaging of the estimates \citep{Geer2016}. Third, it may lead to a non-Gaussian distribution of the forecast error. Previous studies have shown that approximating non-Gaussian error distributions as Gaussian when quantifying the influence of observations using the entropy reduction and the degrees of freedom for signal (DFS) can potentially lead to significantly erroneous estimates \citep{Fowler2013,Fowler2012}.

Convection-permitting DA often uses remote-sensing observations related to clouds and hydrometeors, such as radar reflectivities and cloud-affected satellite radiances, which are nonlinearly related to model state variables. Therefore, the observation operator (sometimes called the observation forward operator/model), used to calculate the model equivalent of observations from a model state, is potentially more nonlinear in convection-permitting NWP \citep[e.g.,][]{Johnson2023,Geiss2021,Scheck2018,SCHECK2016,Kostka2014}. Thus, we need methods that can account for the nonlinearity in the observation operator when assessing the value of observations.

\subsection{Limited area model domain}\label{sec:limited domain}

Convection-permitting NWP usually uses a limited area model \citep[LAM;][]{Gustafsson2018,Dance2004}, which also increases the difficulty of achieving statistically significant estimates of the value of observations. With a global model, we have different weather types happening in different regions at the same time. Therefore, we can easily average the estimates of the value of observations over different weather types. However, with a LAM, we usually only have one weather type at one time, which means we need to run the model for a longer period of time to collect enough samples of different weather types. In convection-permitting NWP, where the focus is often more on accurately forecasting high-impact weather than for global NWP, there is the additional challenge that in order to assess the value of observations on a particular weather type, we need to average over cases of that weather. However, high-impact weather events usually occur at very low frequencies, making it difficult to gather a sufficient number of events of interest, especially over a LAM. Another point related to the domain size and lateral boundary conditions is that after a certain length of time, the observation information will be advected out of the model domain, so this limits the forecast lead-time that observations can have impact for. The short forecast-lead time also causes some problems with using analyses for verification, as for short forecasts the analysis is strongly correlated with the forecast (see section \ref{sec:verification reference} for more discussion). In addition, computing observation bias correction is more difficult in LAMs than in global models \citep{Randriamampianina2005}. This affects the impact that observations can have.

\subsection{New observation types}\label{sec:new observation types}

Convection-permitting NWP is subject to the lack of high-resolution observations in the atmospheric boundary layer. Studies have identified wind profile, temperature profile, humidity profile, precipitation, snow water equivalent and soil moisture as variables currently not adequately measured \citep{Teixeira2021,Leuenberger2020}. It should also be noted that the breakdown of the geostrophic balance at convective scales enhances the need for wind observations because we can no longer infer wind increments from pressure gradients \citep[][]{Bannister2021,Vetra2012}. 

New observing systems have been suggested to fill the observation gap, including ground-based profiling networks \citep{Nomokonova2022,Chipilski2022,PECAN2017}, UAS, high-resolution (geostationary) satellite \citep{waller2016SEVIRI,Scheck2020} and a network collecting crowdsourced observations \citep{bell2022,Hintz2019}, These non-conventional or opportunistic observations usually have a high spatial and temporal resolution and can therefore provide information at appropriate scales \citepalias{WMO-OSCAR}. However, these observations often contain complicated observation errors, such as observation-operator errors \citep{Janjic2018}, spatially and temporally correlated errors \citep{Zeng2021,waller2019DopplerRadarRadialWinds,Michel2018SEVIRI,Cordoba17AMV,waller2016DopplerRadarRadialWinds,waller2016SEVIRI}, and interchannel correlated errors \citep{Stewart2014}. It is important that observation error statistics are accurately specified in DA. For example, the inclusion of spatially correlated error statistics allows observations to provide information on small scales \citep{Fowler2018,stewart2013data,Stewart2008}, and this has been found to improve analysis quality and forecast skill in convection-permitting NWP \citep{Yeh2022,Fujita2020WMO,Simonin2019}.

Before or during the design, development and deployment of an observing system, we need to use a simulation of the observing system to assess its value. Simulating high-resolution observations with complicated errors is more challenging than simulating low-resolution observations with errors whose statistics can be accurately described by just their variance (see section \ref{sec:simulation of observations} for more information).

\section{Available methods for assessing the value of observations in convection-permitting NWP}\label{sec:available methods}

In this section, we briefly introduce existing methods for assessing the value of observations in convection-permitting NWP, including those widely used for global NWP but also applicable to convection-permitting NWP (albeit with some additional considerations) and those recently developed specifically for convection-permitting NWP. We classify the methods into three main categories: 
\begin{enumerate}
    \item methods for quantifying observation influence on the analysis (section \ref{sec:observation influence methods});
    \item methods for quantifying observation impact on forecast skill (section \ref{sec:observation impact methods});
    \item methods for quantifying observation impact on ensemble forecast spread (section \ref{sec:quantifying observation impact on forecast spread}).
\end{enumerate}

In addition to providing guidance on the design of an observation network, these methods may also provide useful information on how DA systems should be improved in order to better assimilate observations. In section \ref{sec:Identification of problems in assimilation systems}, we introduce three methods that have been specifically proposed to identify problems with the assimilation of observations. These methods are an extension of those described in section \ref{sec:observation influence methods} and section \ref{sec:observation impact methods}.

\subsection{Methods for quantifying observation influence on the analysis}\label{sec:observation influence methods}

We first introduce methods for assessing the influence of observations on the analysis, including widely used information content methods and a recently developed method, the partial analysis increment \citep[PAI;][]{Diefenbach2022}. We note that most influence methods measure how much the analysis is \textit{changed} due to the assimilation of observations, not the \textit{accuracy} of the analysis. Furthermore, a larger influence of observations on the analysis does not necessarily imply a larger impact of observations on forecasts. This means that we may need to assess the value of observations on the analysis and forecast separately.

\subsubsection{Information content methods}

Information content methods are used to measure how much information an analysis has retrieved from observations. The same value of information content can be retrieved from a single observation with very high accuracy or from several observations with lower accuracy \citep{RODGERS1998}. A simple way to measure the information content is to calculate the analysis error variance reduction,
\begin{equation}\label{eq:VR}
  \delta\sigma^2=\e{tr}\left(\m{A}\right)-\e{tr}\left(\m{B}\right)
\end{equation}
where $\m{A}\in \reals^{n \times n}$ and $\m{B}\in \reals^{n \times n}$ are the analysis and background error covariance matrices, respectively, and $\e{tr}(\cdot)$ denotes the trace of a matrix. A larger $\delta\sigma^2$ indicates that the analysis has retrieved more information from the assimilated observations, and thus these observations have a larger influence on the assimilation system. Eq. \eqref{eq:VR} can be estimated in observation space \citep{Simonin2019},
\begin{equation}\label{eq:VR in observation space}
    \delta\widetilde{\sigma}^2=\expt{\left(\m{d}^o_a\right)^\top\m{d}^o_a-\left(\m{d}^o_b\right)^\top\m{d}^o_b},
\end{equation}
where 
\begin{equation}\label{eq:residual vector}
    \m{d}^o_a=\m{y}-H(\m{x}_a)
\end{equation}
is the observation-minus-analysis residual vector and 
\begin{equation}\label{eq:innovation vector}
    \m{d}^o_b=\m{y}-H(\m{x}_b),
\end{equation} 
observation-minus-background innovation vector, with $\m{y} \in\reals^m$ the observation vector, $\m{x}_a \in \reals^n$ the analysis state vector, $\m{x}_b \in \reals^n$ the background state vector and $H: \reals^n \rightarrow \reals^m$ the nonlinear observation operator. The symbol $\expt{\cdot}$ denotes the expected value. Applying the observation weighting matrix to Eq. \eqref{eq:VR in observation space}, we obtain \citep{Todling2013}
\begin{equation}\label{eq:Todling2013}
    \delta\widehat{\sigma}^2=\expt{\left(\m{d}^o_a\right)^\top\m{R}^{-1}\m{d}^o_a-\left(\m{d}^o_b\right)^\top\m{R}^{-1}\m{d}^o_b},
\end{equation}
where $\m{R} \in\reals^{m \times m}$ is the observation error covariance matrix and $\m{R}^{-1} \in\reals^{m \times m}$ is its inverse. The expected value in Eqs. \eqref{eq:VR in observation space} and \eqref{eq:Todling2013} is estimated by averaging over a sample of $\m{d}^o_a$ and $\m{d}^o_b$ obtained from cycling the DA system.

The calculation of $\delta \sigma^2$ neglects the non-diagonal elements of the matrices. An alternative method that includes the off-diagonal elements is the reduction in entropy \citep[][section 2.2]{Shannon1949} of the error covariance matrices \citep[][]{Fisher2003,Rodgers2000,RODGERS1998},
\begin{equation}\label{eq:entropy reduction}
    \delta S=\frac{1}{2}\e{log}_2\lvert\m{A}\rvert-\frac{1}{2}\e{log}_2\lvert\m{B}\rvert,
\end{equation}
where $\lvert \cdot \rvert$ denotes the determinant of a matrix. The entropy reduction is sometimes referred to as mutual information \citep{Fowler2017,Fowler2013,Cover1991}. In variational DA, Eq. \eqref{eq:entropy reduction} can be simplified by using a control variable transform \citep{Fisher2003}. How $\delta S$ differs from $\delta \sigma^2$ when observations have correlated errors has been discussed in \citet{Fowler2019compression}. An important use of $\delta S$ is for the channel selection of hyperspectral satellite instruments \citep{Fowler2017,Rodgers2000}. 

Another method that belongs to the class of information content methods is the DFS, which will be described separately in the following section due to its widespread use and a large amount of relevant research \citep[e.g.,][]{Hotta2021,Fowler2020,Desroziers2009,Chapnik2006,Desroziers2005,Cardinali2004,Fisher2003,Desroziers2001}. 

\subsubsection{Degrees of freedom for signal (DFS)}\label{sec:DFS}

The name of the DFS comes from the idea that the model state can be spanned by $n$ orthogonal vectors, and hence it can vary statistically independently in $n$ directions \citep[``degrees of freedom";][section 2.4]{Rodgers2000}. If the error uncertainty in one direction is constrained well by the observations, then it can be considered to represent one ``degree of freedom for signal". Directions that are not well constrained represent the ``degrees of freedom for noise". Therefore, a larger DFS indicates a larger influence of the observations on the analysis. The DFS can be formulated as \citep{Fowler2020,Lupu2011}
\begin{equation}\label{eq:DFS HK}
    \e{DFS}=\e{tr}\left(\m{HK}\right),
\end{equation}
where $\m{H} \in\reals^{m \times n}$ is the linearized observation operator and $\m{K}\in\reals^{n\times m}$ the Kalman gain matrix \citep{Nichols2010,Bouttier2002}. There is more than one way to express the DFS \citep{Fowler2020,Chapnik2006,Cardinali2004,Fisher2003}, but they can all lead to Eq. \eqref{eq:DFS HK}. For different formulations for the DFS see Appendix \ref{appendix:DFS A}.

Eq. \eqref{eq:DFS HK} cannot really be used in practice to calculate the DFS for several reasons: 1) the Kalman gain matrix is usually not formed explicitly in operational DA systems; 2) the observation operator can be nonlinear; and 3) the calculation of the trace of large matrices is expensive. For information on how the DFS is calculated in practice see Appendix \ref{appendix:DFS B}.

\subsubsection{Partial analysis increments (PAIs)}\label{sec:partial analysis increments}

\citet{Diefenbach2022} proposed to use the PAIs as a 3-D observation influence measure and a diagnostic for ensemble data assimilation systems. In ensemble-based DA, the analysis increment vector can be calculated as 
\begin{equation}\label{eq:analysis increment}
    \delta \m{x}=\m{K}\m{d}^o_b,
\end{equation}
where the Kalman gain matrix is given by \citep{Sommer2016}
\begin{equation}\label{eq:K ensemble estimate}
    \m{K}=\frac{1}{N-1}\m{X}_a\left(\m{Y}_a\right)^\top\m{R}^{-1},
\end{equation}
with $N$ being the ensemble size, $\m{X}_a \in \reals^{n \times N}$ the analysis ensemble perturbation matrix whose $i$-th column is the difference of the $i$-th ensemble member from the ensemble mean, $\m{Y}_a \in \reals^{m \times N}$ the analysis ensemble perturbation matrix in observation space whose $i$-th column is the difference of the $i$-th ensemble member of $H(\m{x}_a)$ from the ensemble mean of $H(\m{x}_a)$ \citep[see][]{Livings2008}, and $\m{R} \in\reals^{m \times m}$ the observation error covariance matrix. It should be noted that Eq. \eqref{eq:K ensemble estimate} is only exact in the absence of localization. However, according to \citet{Diefenbach2022}, applying localization to Eq. \eqref{eq:K ensemble estimate} only introduces minor errors.

Eq. \eqref{eq:analysis increment} shows that the analysis increment for each state variable is determined by a weighted sum of the elements of the vector $\m{d}^o_b$ (with the weights given by the elements of the Kalman gain, $\m{K}$). Then, the PAI 
 gives the influence of an observation $j$ on the state variable $i$ as  
\begin{equation}\label{eq:PAIs}
    \left[\delta x^{(j)}\right]_i=K_{i,j}\left[d^o_b\right]_j,
\end{equation}
where $K_{i,j}$ is the element of the matrix $\m{K}$ in the $i$-th row and $j$-th column, and $\left[d^o_b\right]_j$ the $j$-th element of the vector $\m{d}^o_b$. We call $\delta \m{x}^{(j)}\in\reals^n$ the PAI vector and  $\left[\delta x^{(j)}\right]_i$ denotes its $i$-th element.  The sum of PAIs for all observations is the analysis increment vector, i.e., $\left[\delta x\right]_i = \sum_{j=1}^m\left[\delta x^{(j)}\right]_i$. The PAI vector describes the 3-D influence of an observation in model space. This allows the PAI to be used to assess the influence of observations on each model state variable, which can be useful to optimize the localization used in the assimilation. The relative magnitude of the PAI vector represents the overall influence of an observation. In addition, the statistics of the PAIs (e.g., mean, standard deviation and absolute mean) can be used to examine the systematic increments resulting from a certain type of observations \citep{Diefenbach2022}. Furthermore, by detecting whether different observations draw the analysis in opposite directions, the PAIs may be used to identify the suboptimality in the assimilation of observations \citep{Diefenbach2022}.

It should be noted that the PAIs resulting from a single observation differ from the analysis increment in a single-observation experiment \citep{Diefenbach2022}. The PAIs represent the influence an observation has when it is assimilated with other observations. In contrast, the analysis increment in a single-observation experiment represents the influence an observation has when it is assimilated alone. If we compare the PAI method with the DFS, a difference between them is that the PAI method assesses the influence of observations on the first moment of the posterior estimate of the model state (mean analysis), while the DFS assesses the influence of observations on the second moment of the posterior (analysis variance).

\subsection{Methods for quantifying observation impact on forecast skill}\label{sec:observation impact methods}

We now introduce methods for assessing the impact of observations on forecast skill. Due to the stronger nonlinearity of the error growth, the forecast lead-time of interest in observation impact studies for convection-permitting NWP is much shorter than that in global observation impact studies. The methods introduced in this section may also be used to assess the influence of observations on the analysis by setting the forecast lead-time to zero and using observations as the verification reference.

\subsubsection{Observing system experiments/Data denial experiments (OSEs/DDEs)}\label{sec:OSE}

An observing system experiment \citep[OSE;][]{Bouttier2001,Cress2001} is a common tool that has been used for decades in operational NWP centres to assess the impact of observations in an operational NWP system. An OSE is also known as a data denial experiment (DDE) because it is carried out by first running the NWP system, assimilating all observations currently in use, and then running the system without assimilating the subset of observations for which we want to derive the impact of \citep[e.g.,][]{Candy2021,Eyre2021a,Gelaro2009}. The impact of the removed observations is assessed by the difference in forecast skill between the two runs. Forecast skill is measured by comparing the forecasts to a proxy for the truth, which is typically a verifying analysis or observations \citep[e.g.,][]{Lawrence2019}. Because an OSE assesses the actual improvement (or degradation) in forecast skill of operational NWP, OSEs can serve as a standard for validating the results of other methods that may use synthetic observations (e.g., an observing system simulation experiment [OSSE]), assume a linear growth in the forecast error (e.g., the ensemble forecast sensitivity to observation impact [EFSOI]), or use a proxy for forecast skill (e.g., ensemble sensitivity analysis [ESA]). An OSE can provide insight into how current observing systems should evolve, but it cannot actually measure the impact of future observing systems. Many studies have used OSEs to study the impact of observations in the context of convection-permitting NWP \citep[e.g.,][]{Johnson2022,Green2022,Chipilski2022,Chipilski2020,Caumont2016}.

\subsubsection{Observing system simulation experiments (OSSEs)}\label{sec:OSSE}

An OSSE is used to assess the impact of a \textit{future} observing system \citep[e.g.,][]{Hoffman2016,Errico2018}. An OSSE is similar to an OSE, except that the OSSE uses synthetic observations and a hypothetical true state that is given by a ``nature run" (NR) from an NWP model (e.g., running a more sophisticated model than the model used to assimilate the data for several months). This nature run is used to generate synthetic observations and validate forecasts. Synthetic observations are generated by interpolation of the NR fields and sometimes the accompanying use of an observation operator, such as a radiative transfer model. Simulated observation errors should be added to the synthetic observations.  The use of synthetic observations allows an OSSE to offer greater flexibility than an OSE. In addition to assessing the impact of a future observation type, an OSSE may be used to assess the impact of an alternate deployment of an existing observation type \citep[e.g., different spatial and temporal resolutions and geophysical locations;][]{Prive2022}.  A few studies have used OSSEs to assess the impact of observations in the context of convection-permitting NWP \citep[e.g.,][]{Huang2022,Maejima2022}. 

In order to achieve reliable results, OSSE systems should keep pace with the development of operational NWP systems and real observing systems. However, an important caveat of OSSEs is that the results are dependent on current DA capabilities, and do not provide a projection of the capabilities of future versions of DA (including scientific, technical, and computing improvements that may be in place when a new observing system becomes operational). Furthermore, OSSE systems should be calibrated (e.g., using the corresponding OSE) to ensure that their results are comparable to those that would be obtained using real observing systems.

\subsubsection{Sensitivity observation system experiments (SOSEs)}\label{sec:SOSE}

The sensitivity observation system experiment (SOSE) developed by \citet{Marseille2008} is another method for assessing the impact of a \textit{future} observing system. In contrast to an OSSE, the baseline of the observing system in a SOSE is given by real observations, and only future observations are simulated in a SOSE. Another difference is that synthetic observations in a SOSE are generated using an adapted analysis rather than the nature run. The adapted analysis is created by the following steps: 1) run a forecast from the background, 2) compute the forecast error with respect to a verifying analysis, 3) project the forecast error back to the initial time through a forecast sensitivity computation \citep{Rabier1996}, and use the resulting key background errors to correct the background, and 4) create a new analysis using the corrected background and all existing observations. Readers are referred to \citet[][Figure 1]{Marseille2008} for a schematic diagram of the SOSE analysis scheme. Since the SOSE is not as widely used as the OSSE, it is not yet clear how well the SOSE performs in convection-permitting NWP.

\subsubsection{Ensemble forecast sensitivity to observation impact (EFSOI)}\label{sec:EFSOI}

The EFSOI \citep{Kalnay2012,Li2010,Liu2008} was developed based on the forecast sensitivity to observation impact (FSOI) originally proposed by \citet{Langland2004}. These methods measure how much a subset of observations contributes to the reduction of short-range forecast error within a system assimilating all observations \citep{Eyre2021b}. Observations that contribute to a greater reduction could be considered to have a greater impact. The reduction of the forecast error due to the assimilation of observations is measured by \citep{Fourrie2002}
\begin{equation}\label{eq:delta e}
    \delta e = \left(\m{x}^f_a-\m{x}_t\right)^\top\m{C}\left(\m{x}^f_a-\m{x}_t\right)-\left(\m{x}^f_b-\m{x}_t\right)^\top\m{C}\left(\m{x}^f_b-\m{x}_t\right),
\end{equation}
where $\m{x}^f_a \in\reals^n$ is a vector of the forecast generated from the analysis, $\m{x}^f_b\in\reals^n$ a vector of the forecast generated from the background, $\m{x}_t \in\reals^n$ a vector of the verifying analysis and $\m{C} \in\reals^{n\times n}$ a symmetric matrix of energy weighting coefficients often representing total energy \citep{Ehrendorfer1999,rosmond1997technical,Rabier1996,Ehrendorfer1995}. The choice of the matrix $\m{C}$ affects the calculated value of forecast error reduction and thus may have a large effect on the estimation of observation impacts \citep{Janiskova2016}. The EFSOI is then given by an approximation of $\delta e$ (see Appendix \ref{appendix:EFSOI derivation}),
\begin{equation}\label{eq:EFSOI}
    \delta e \approx \e{EFSOI} = \frac{1}{N-1}\left(\overline{\m{d}}^o_b\right)^\top\m{R}^{-1}\m{Y}_a\left(\m{X}_a^f\right)^\top\m{C}\left[\left(\overline{\m{x}}^f_a-\m{x}_t\right)+\left(\overline{\m{x}}^f_b-\m{x}_t\right)\right],
\end{equation}
where the overline denotes the ensemble mean and $\m{X}^f_a \in\reals^{n\times N}$ is the analysis forecast ensemble perturbation matrix, which is obtained in a similar way as the matrix $\m{X}_a$ (see section \ref{sec:partial analysis increments}), but using a forecast ensemble. We may consider 
 Eq. \eqref{eq:EFSOI} as a product of $\left(\overline{\m{d}}^o_b\right)^\top$ and a 
sensitivity vector describing how small changes in observations will change the forecast error reduction, 
\begin{equation}
    \left[\frac{\partial\delta e}{\partial \m{y}}\right] = \frac{1}{N-1}\m{R}^{-1}\m{Y}_a\left(\m{X}_a^f\right)^\top\m{C}\left[\left(\overline{\m{x}}^f_a-\m{x}_t\right)+\left(\overline{\m{x}}^f_b-\m{x}_t\right)\right].
\end{equation}
The FSOI and EFSOI are very similar in concept. A major difference is that in the FSOI the sensitivity vector is computed from the adjoint of the tangent linear model of the forecast model (see Appendix \ref{appendix:EFSOI derivation}). We focus on the EFSOI because of a lack of adjoint models useful at convective scale and because the EFSOI is more computationally robust to the strong nonlinearity in the forecast error growth. 

In Eq. \eqref{eq:delta e}, the analysis is used to verify the forecast. If observations are used, we obtain the observation-based EFSOI whose sensitivity vector has a different expression than the EFSOI with self-analysis verification (see Appendix \ref{appendix:observation-based verification}). We discuss whether the analysis or observations should be used to verify the forecast in convection-permitting NWP in section \ref{sec:verification reference}. 

The EFSOI has been used in convection-permitting NWP as a computationally inexpensive alternative to the OSE/DDE \citep[e.g.,][]{Gasperoni2022,Necker2018}. The EFSOI has also been used to identify the problems with the assimilation system \citep{Ota2013} and to provide a fully flow-dependent quality control \citep{Chen2020,Chen2019,Hotta2017}.

\subsubsection{Comparison between the EFSOI and OSE/DDE}

Several differences between the EFSOI and DDE should be noted when comparing their results \citep{Eyre2021a}. First, the EFSOI uses a linear approximation of forecast error growth, whereas a DDE allows for fully nonlinear forecast error growth. Second, the EFSOI is calculated using the error covariance matrices of the full observing system only, whereas a DDE consists of two forecast runs, one using the error covariance matrices of the full observing system and the other using adjusted error covariance matrices as some observations are removed. In practice, the differences in treatment of covariance localization between the two approaches will also contribute to differences in their results. Third, the EFSOI measures non-cycled observation impacts, whilst a DDE can be used to assess the accumulated impact of observations in a cycled DA environment (for more details see section \ref{sec:ability to assess the accumulated value of observations}). Because of these differences, the EFSOI of two observation subsets should be equal to the sum of the EFSOI of each observation subset, which is however not the case for a DDE. 

\subsection{Methods for quantifying observation impact on forecast spread}\label{sec:quantifying observation impact on forecast spread}

In ensemble forecasting, the ensemble spread (e.g., calculated as the standard deviation between ensemble members) is an estimate of the uncertainty in the forecast and should be negatively correlated with forecast skill. Due to this spread-skill relationship, assessing the impact of observations on the ensemble spread is qualitatively equivalent to assessing the impact of observations on forecast accuracy. Specifically, a larger reduction in the ensemble spread indicates a larger improvement in forecast skill. 

In this section, we introduce two methods that assess the impact of observations by measuring the change in ensemble spread. The successful use of these methods is dependent on a clear spread-skill relationship. These methods are advantageous for studying the impact of future observations because they only focus on the impact on the ensemble spread, not the ensemble mean. This means that, unlike an OSSE, simulated observations do not need to be consistent with an underlying truth and so a separate nature run is not needed.

\subsubsection{Spread reduction in ensemble of data assimilations (EDA)}\label{sec:EDA}

Ensemble of data assimilations \citep[EDA;][]{Isaksen2010,Bonavita2016} is an ensemble of independent three-dimensional or four-dimensional data assimilations, which is primarily designed to estimate flow-dependent background and analysis error statistics for the hybrid 4D-Var systems. The EDA may also be used to provide initial conditions for ensemble forecasting. When assessing the value of observations, the EDA allows for a statistical estimate of uncertainty reduction in analyses and short-range forecasts due to the assimilation of additional real or synthetic observations \citep{Tan2007,Harnisch2013}. If the new observations added to the DA system result in a larger reduction in the EDA spread, they are considered to have a greater value.

\subsubsection{Ensemble sensitivity analysis (ESA)}\label{sec:ESA}

The ESA \citep{Ancell2007} provides a computationally cheap way of estimating the reduction in the ensemble variance of a forecast metric due to the assimilation of observations \citep{Griewank2023,Nomokonova2022}. Let $J=f(\m{x})$ be a scalar function of model state variables, quantifying aspects of the forecasting system of interest, and $\m{J} \in \reals^{N}$ be a vector containing a set of $J$ computed for each ensemble member, then the reduction in the ensemble variance of $J$ is calculated as 
\begin{equation}\label{eq:EVR of J}
    \delta\sigma^2_J=\left[\frac{\partial J}{\partial \m{x}}\right]^\top(\m{P}_a-\m{P}_b)\left[\frac{\partial J}{\partial \m{x}}\right],
\end{equation}
where $\left[\partial J/\partial \m{x}\right]\in \reals^n$ is a sensitivity vector that describes how small changes in model state variables at the initial time will change $J$ at a forecast lead-time,
\begin{equation}\label{eq:P_a}
    \m{P}_a=\frac{1}{N-1}\m{X}_a\left(\m{X}_a\right)^\top
\end{equation}
an ensemble estimate of $\m{A}$ and 
\begin{equation}\label{eq:P_b}
    \m{P}_b=\frac{1}{N-1}\m{X}_b\left(\m{X}_b\right)^\top
\end{equation}
an ensemble estimate of $\m{B}$, with $\m{X}_b \in \reals^{n \times N}$ the background ensemble perturbation matrix. For a derivation of Eq. \eqref{eq:EVR of J} see Appendix \ref{appendix:ESA derivation}.

The value of $\delta\sigma^2_J$ can be calculated without explicitly knowing the sensitivity vector (see Appendix \ref{appendix:ESA derivation implicit}). However, this implicit calculation does not allow the covariance localization used in the assimilation to be easily included in the computation of $\delta\sigma^2_J$. If the sensitivity vector is calculated explicitly (see Appendix \ref{appendix:ESA derivation explicit}), then $\delta\sigma^2_J$ can be calculated using Eq. \eqref{eq:EVR of J}, and the covariance localization can be easily included. For a more detailed discussion of the localization in the ESA method, the reader is referred to \citet{Griewank2023}. Assuming the sensitivity vector and the background ensemble perturbation matrix are available, the calculation of Eq. \eqref{eq:EVR of J} only requires the analysis ensemble perturbation matrix, which can be obtained if we know the error statistics of the observations to be assessed and have the corresponding observation operator (for more details see Appendix \ref{appendix:ESA derivation explicit}). Without the need to know observation values, the ESA method is particularly well-placed to assess the impact of future observations.

\subsubsection{Comparison between the ESA method and EFSOI}

The major difference between the ESA method and the EFSOI is that the ESA method assesses the impact of changes in the ensemble perturbation matrix, $\m{X}_a - \m{X}_b$, on forecast spread, whereas the EFSOI can be seen as assessing the impact of the analysis increment, $\m{x}_a - \m{x}_b$, on forecast skill. Moreover, the EFSOI uses a fixed scalar forecast error metric, whereas the scalar forecast metric, $J$, used in the ESA method is an arbitrary function of the forecast state, and a different $J$ can be used at different locations \citep{Tardif2022}.  This makes the ESA method attractive for specific applications (e.g., quantitative precipitation forecasts or forecasting for renewable energy applications). Another difference is that the ESA method does not require knowledge of the observation value ($\m{y}$), whereas the EFSOI does.

\subsubsection{Comparison between the ESA and EDA methods}

There are two main differences between the ESA and EDA methods. First, the EDA method calculates the reduction in the standard deviation (or other spread measures) of the ensemble of model state variables, while the ESA method calculates the reduction in the ensemble variance of $J$, which is a scalar function of model state variables. Second, the EDA method allows for a fully nonlinear growth in the ensemble spread of model state variables, whereas the ESA method uses a linear approximation to calculate the ensemble variance of $J$ caused by the initial ensemble perturbations of model state variables (i.e., the use of the sensitivity vector).

\subsection{Identification of problems in assimilation systems}\label{sec:Identification of problems in assimilation systems}

How well observations are assimilated affects their value for NWP. For example, if background and observation error statistics are incorrectly specified in the assimilation (i.e., the error covariance matrices used in the assimilation do not accurately represent the true error statistics), or if the observation operator is inaccurate, then the observations may not be optimally assimilated and thus their values may not be correctly represented. In this section, we discuss how to identify issues related to the assimilation of observations using methods based on the DFS and EFSOI. We note that methods described in sections \ref{sec:observation influence methods} - \ref{sec:quantifying observation impact on forecast spread} may also be used to identify problems in the assimilation system.

\subsubsection{The actual value of the DFS in suboptimal DA systems} 

\citet{Fowler2020} have shown that the value of DFS calculated under the assumption of optimality of a DA system (Eq. \eqref{eq:DFS HK}) may differ from the actual value of the DFS in a suboptimal system, and this difference can be used to identify the suboptimality in the assimilation of observations.

Let $\m{D}=\m{H}\m{B}\m{H}^\top+\m{R}$ denote the innovation covariance matrix given by the background and observation error covariance matrices used in the assimilation and $\m{D}_t=\expt{\m{d}^o_b\left(\m{d}^o_b\right)^\top}$ denote the true innovation covariance matrix. If $\m{D} \neq \m{D}_t$, then Eq. \eqref{eq:DFS HK} does not hold, and the actual DFS should be calculated as \citep{Fowler2020}
\begin{equation}\label{eq:DFS actual}
    \e{DFS}_{Actual}=\e{tr}\left(\m{H}\m{K}\m{D}_t\m{D}^{-1}\right).
\end{equation}
In practice, the mismatch between $\m{D}$ and $\m{D}_t$ can be caused by 1) misspecification of the background and observation error statistics in the assimilation, 2) linearization of a nonlinear observation operator, or 3) mutual correlations between observation errors and background errors \citep{Fowler2020,Lupu2011}. 

A comparison between the theoretical DFS given by Eq. \eqref{eq:DFS HK} and the actual DFS given by Eq. \eqref{eq:DFS actual} allows us to find observations that are
not optimally assimilated and are potentially damaging to the analysis. \citet{Fowler2020} calculated the two values of the DFS using assimilation residuals (see Appendix \ref{appendix:DFS B}) and found in the Met Office's convection-permitting NWP system that poor assumptions in the observation operator led to the actual DFS up to three times larger than its theoretical value when assimilating Doppler radial winds. 

This discrepancy in the actual and theoretical degrees of freedom has been used to tune background and observation error covariance matrices with some success \citep{Chapnik2006,Desroziers2001}. However, as noted in \citet{Fowler2020} the source of the discrepancy cannot be uniquely diagnosed from the difference between the actual and theoretical DFS alone.

\subsubsection{Cross-validation (C-V) of observations}\label{sec:cross-validation}

\citet{Stiller2022} proposed new diagnostics to test the suboptimality in the assimilation of observations in ensemble-based DA systems based on partitioning the EFSOI into two terms. The first term is the cross-validation (C-V) term, which validates whether the assimilation of an observation type (observations to assess) pulls the model state towards the verifying observations (e.g., another observation type known to have good quality; see Appendix \ref{appendix:C-V diagnostic 1}). The remaining term is proportional to the analysis increments and in an optimal assimilation system should be equal to the C-V term, thus allowing the optimality of the assumptions made to be tested (see Appendix \ref{appendix:C-V diagnostic 2}).

Calculating the C-V term involves the product of innovation vectors of two independent observation types (Eq. \eqref{eq:C-V Jb trace}. As explained in Appendix \ref{appendix:C-V diagnostic 1} (under assumptions usually made in DA), the statistical expectation of such products yields background error covariances in observation space (when verification and assimilation times are identical). Alternatively, background error covariances can be estimated using ensemble perturbation matrices (Eq. \eqref{eq:ensemble estimator}). \citet{Stiller2022} has shown that computing the C-V term separately using these two approximations to the background error covariances gives good agreement when calculated from two trusted observation types (e.g., in-situ observations). The result obtained from two trusted observation types can then be used as a benchmark for identifying observation-related issues when cross-validating more problematic observations. As a different application, results from two trusted observation types can also be used to test an ensemble estimate of the background error covariance matrix (compare Eq. \eqref{eq:C-V diagnose 1 consistency}) or to assess different choices of localization parameters.

\citet{Stiller2022} has also developed a single-observation version of the C-V diagnostics, where the diagnostic values are calculated from a large number of assimilation processes, each assimilating only a single observation. Showing similar results regarding the suboptimality in the assimilation of observations, the single-observation version is easier and more flexible to apply. The single-observation version is particularly suitable for testing observations that are not yet routinely used in the assimilation.

Since the proposed diagnostics are statistical quantities, to facilitate their interpretation, \citet{Stiller2022} also presented an indicator of the statistical significance for the C-V term \citep[][Appendix C]{Stiller2022}, as well as a normalization that can make the statistical comparison largely independent of the total number of data and the closeness of their collocation.

\subsubsection{Ensemble forecast sensitivity to observation error covariance (EFSR)} 

The ensemble forecast sensitivity to the matrix $\m{R}$ (EFSR) was proposed by \citet{Hotta2017EFSR} based on the EFSOI and the forecast sensitivity to observation error covariance derived by \citet{Daescu2008}. The EFSR describes how the forecast error changes due to small changes in the observation error covariance matrix,
\begin{equation}\label{eq:FSR}
    \frac{\partial e_a}{\partial R_{i,j}}=-\left[\m{R}^{-1}\m{d}^o_a\right]_j\left[\frac{\partial e_a}{\partial \m{y}}\right]_i,
\end{equation}
where $R_{i,j}$ is the element of the matrix $\m{R}$ in the $i$-th row and $j$-th column, $\left[\m{R}^{-1}\m{d}^o_a\right]_j$ the $j$-th element of the product of $\m{R}^{-1}$ and $\m{d}^o_a$, and $\left[\partial e_a/\partial \m{y}\right]_i$ the $i$-th element of the sensitivity vector, $\left[\partial e_a/\partial \m{y}\right]$. The forecast error is measured by (see Eq. \eqref{eq:delta e})
\begin{equation}\label{eq:e_a}
    e_a=\left(\m{x}^f_a-\m{x}_t\right)^\top\m{C}\left(\m{x}^f_a-\m{x}_t\right).
\end{equation}
Using the chain rule for partial derivatives, changes in forecast error due to small changes in the observation values are described by \citep{Hotta2017EFSR} 
\begin{equation}\label{eq:EFSR}
   \left[\frac{\partial e_a}{\partial \m{y}}\right]=\frac{2}{N-1}\cdot\m{R}^{-1}\m{Y}_a\left(\m{X}^f_a\right)^\top\m{C}\left(\overline{\m{x}}^f_a-\m{x}_t\right).
\end{equation}
The EFSR can be used to diagnose whether the prescribed observation error variance is optimal. A positive value of $\partial e_a/\partial R_{i,j}$ means that $e_a$ increases with $R_{i,j}$. In this case, we should decrease the observation error variance to reduce the forecast error. In contrast, if $\partial e_a/\partial R_{i,j}$ is negative, we should increase the observation error variance to reduce the forecast error.

\section{Advantages and disadvantages of available methods}\label{sec:pros and cons}

In this section, we compare the advantages and disadvantages of different methods in terms of their ability to address the challenges described in section \ref{sec:Challenges in C-P NWP}. We summarize the comparison in Table \ref{tab:features}, listing the features that can be considered advantages and indicating which methods have which features. Each feature is discussed in more detail in the following subsections.

\begin{table}
    \centering
    \begin{tabular}{|m{9em}|c|c|c|c|c|c|c|c|c|c|}\hline
        \diagbox[width=10.2em]{Features}{Methods}& 
        DFS   &PAI  &OSE  &OSSE &SOSE &EFSOI &EFSR &C-V  &ESA  &EDA    \\ 
        \hline 
        Applicable for ensemble forecasting systems 
        &$\y$ &$\y$ &$\y$ &$\y$ &$\s$ &$\y$  &$\y$ &$\y$ &$\y$ &$\y$   \\
        \hline
        Targetable via user-specified metric
        &--   &--   &$\y$ &$\y$ &$\y$ &$\y$  &$\y$ &$\y$ &$\y$ &$\n$   \\
        \hline 
        Applicable for future observations     
        &$\s$ &$\s$ &$\n$ &$\y$ &$\y$ &$\s$  &$\s$ &$\s$ &$\y$ &$\y$   \\
        \hline 
        Computationally inexpensive 
        &$\y$ &$\y$ &$\n$ &$\n$ &$\n$ &$\y$  &$\y$ &$\y$ &$\y$ &$\s$   \\   
        \hline 
        Able to consider the nonlinearity of observation operators
        &$\s$ &$\s$ &$\y$ &$\y$ &$\y$ &$\s$  &$\s$ &$\s$ &$\s$ &$\y$   \\
        \hline
        Without the assumption of linear error growth
        &--   &--   &$\y$ &$\y$ &$\s$ &$\n$  &$\n$ &$\n$ &$\n$ &$\y$   \\
        \hline 
        Able to obtain contributions from individual observations
        &$\y$ &$\y$ &$\n$ &$\n$ &$\n$ &$\y$  &$\y$ &$\y$ &$\y$ &$\n$   \\
        \hline
        Able to assess the accumulated value of observations 
        &$\n$ &$\n$ &$\y$ &$\y$ &$\s$ &$\n$  &--   &--   &$\n$ &$\y$   \\
        \hline
        Able to assess the value of anchoring 
        &$\n$ &$\y$ &$\y$ &$\y$ &$\s$ &$\y$  &--   &--   &$\n$ &$\n$   \\
        \hline
    \end{tabular}
    \caption{Features of different methods can be considered as advantages when assessing the value of observations in convection-permitting NWP. The ability to be used in ensemble forecasting systems is treated as an advantage because such systems are often used for convection-permitting NWP. The symbol $\y$ means ``yes'', the symbol $\n$ means ``no'', the symbol $\s$ means somewhere in between and the symbol ``--'' means ``not applicable''.} \label{tab:features}
\end{table}

\subsection{Applicability for ensemble forecasting systems}\label{sec:suitable for ensemble forecasting systems}

The first step in choosing a suitable method is to check whether that method can be used in a given NWP system. Since many current operational convection-permitting NWP systems are either already operating ensemble forecasting systems or are moving toward this direction, it is important to have suitable methods that can be used for ensemble forecasting systems. Generally speaking, all the methods listed in Table \ref{tab:features} are applicable in principle to such systems. However, some methods may require further research. For example, the PAI and EFSOI were originally developed for the LETKF system \citep{HUNT2007}. Conceptually, these methods should work for other ensemble-based DA systems, but there may be significant practical issues to overcome. In addition, calculating the DFS for ensemble-based DA systems is largely an unexplored area, although we already have some methods that may be applicable to ensemble-based DA (see Appendix \ref{appendix:DFS B}). It is also unclear how to perform a SOSE using an ensemble forecasting system. One idea is to use ensemble perturbation matrices instead of the adjoint model \citep{Marseille2008}. 

\subsection{Targetability via user-specified metric}

 The use of forecast verification metrics in a method allows us to focus on quantifying forecast errors for features relevant to high-impact weather events (e.g., quantitative precipitation forecasting, high winds or fog), or forecasting for particular user requirements (such as renewable energy forecasting). In  Table \ref{tab:features}, we have indicated the methods that are able to use user-specified metrics in their calculations.  

\subsection{Applicability for future observations}\label{sec:Applicability for future observations}

Convection-permitting NWP requires new observation types that may not be currently available. Therefore, to assess the value of these novel observations, we need to use synthetic observations that realistically represent the error statistics of these new observations. By definition, an OSE is only used to assess the value of observations from existing observing systems. In contrast, an OSSE, a SOSE, the EDA method and the ESA method are specifically used to assess the value of future observations. One difference between an OSSE and the other methods is that in an OSSE we simulate all observations (including existing and future observations), while in the other methods, we simulate only future observations (or observation error statistics). A key point in the applicability of the ESA method for assessing the impact of future observations is that it does not require observations (only observation error statistics are required). If the DFS is calculated directly using the matrices, $\m{H}$ and $\m{K}$ (Eq. \eqref{eq:DFS HK}), then the calculation of the DFS also does not require observations. However, in the actual calculations of the DFS, it requires observations.

The EFSOI was developed as a computationally inexpensive alternative to the OSE, so it is usually used with real observations. However, it should be noted that the EFSOI can be calculated using synthetic observations in the context of an OSSE. Such experiments can be used to study how different factors (e.g., forecast length) affect the accuracy of the impact estimated by the EFSOI. For example, \citet{Prive2021} evaluated the FSOI as a function of forecast length in an OSSE. Similarly to the EFSOI, the DFS, PAIs, the EFSR and the C-V diagnostics can be computed in an OSSE. 

\subsection{Computational expense}\label{sec:computational cost}

Computational cost is an important factor to consider when choosing a method to assess the value of observations, especially for convection-permitting NWP where a statistically significant estimate is more difficult to obtain (see sections \ref{sec:nonlinearity} and \ref{sec:limited domain}). The actual computational cost depends on the complexity of the convection-permitting NWP system, the size of the problem (e.g., the size of the model domain and the number of observations) and computational resources (e.g., the power of the supercomputer). Here, we compare the relative computational expenses of different methods. An OSE, an OSSE, or a SOSE generally requires the highest computational cost due to the need to run a full operational NWP system for assessing the impact of a change to the observing system \citep{Eyre2021a}. It is possible to use simpler configurations of the EDA \citep[e.g., lower resolution and fewer members;][]{Lang2019} to assess the value of observations in global NWP. However, given the importance of resolution in convection-permitting NWP, this may not be appropriate for assessing the value of observations in convection-permitting NWP. In contrast, other methods can be calculated at a relatively low cost. The DFS can be calculated using a sample of assimilation residuals (Eq. \eqref{eq:DFS residuals} and Eq. \eqref{eq:DFS actual residuals}), which are by-products of a variational DA system. The PAI method, the EFSOI, the EFSR, the C-V diagnostics and the ESA method make use of the ensemble perturbation matrices produced by an ensemble forecasting system and do not require the system to be rerun when assessing a different set of existing observations (i.e., observations already assimilated in the system). In addition, the ESA method (explicit calculation) requires only the calculation of analysis ensemble perturbations for given observation error statistics when assessing new/future observations.

\subsection{Ability to consider the nonlinearity in observation operators}\label{sec:Applicability for nonlinear observation operators}

When carrying out an OSE or an OSSE, or calculating the EDA spread reduction, we can use linear or nonlinear
observation operators, consistent with the DA algorithm used. The PAI method, the EFSOI, the C-V diagnostics, the EFSR and the ESA method are used in ensemble-based DA systems (e.g., LETKF), which are commonly based on the implicit assumption of a linear observation operator. These methods can match the analysis update in ensemble-based DA systems exactly, and thus they can reflect the actual value of observations in such DA systems. The DFS can be calculated using the output of a DA system that uses nonlinear observation operators, although the practical implementation of the DFS may be derived based on the linear assumption of the observation operator (see Appendix \ref{appendix:DFS B} for details).

\subsection{Absence of the assumption of linear error growth}\label{sec:Absence of the assumption of linear error growth}

A linear approximation of forecast error growth is used in the derivation of the FSOI (Eq. \eqref{eq:linearised forecast model}). The EFSOI is derived from the FSOI, and the EFSOI further assumes a linear relation between initial ensemble perturbations and forecast ensemble perturbations (Eq. \eqref{eq:EFSOI linear assumption}). The EFSR and C-V diagnostics are developed from the EFSOI, and thus they make the same assumptions as the EFSOI. In the ESA method, initial ensemble perturbations of model state variables are linearly projected to the perturbations of the forecast metric, $J$ (Eq. \eqref{eq:ESA sensitivity}). Although these methods rely on a similar linear assumption, they are robust to the strong nonlinear error growth due to the use of ensemble perturbations. As a result, these methods are useful in convection-permitting NWP. In a SOSE, a linear approximation of forecast error growth is used to generate synthetic observations (see section \ref{sec:SOSE}). In general, the violation of the assumption of linear error growth becomes more significant as forecast lead-time increases and scales reduce \citep{Ancell2022}.

\subsection{Ability to obtain contributions from individual observations}

An advantage of assuming a linear forecast error growth in assessing the impact of observations is that the impact estimate can be partitioned into contributions from individual observations, or the contributions from many observations can be aggregated to give an impact estimate for a set of observations \citep{Kalnay2012,Stiller2022, Tardif2022}. Partitioning and aggregation make it easy to assess the value of different subsets of observations (e.g., without the need to re-run the NWP system). When assessing the influence of observations, we can also obtain contributions from individual observations by partitioning the results of the DFS and PAIs \citep{Hotta2021,Diefenbach2022}. 

It should be noted that the value of a single observation estimated by partitioning the value of a set of observations is different from the value of a single observation estimated using a single-observation DA experiment. A single-observation experiment measures the value of a single observation when it is assimilated alone, while partitioning gives the value of a single observation when it is assimilated with other observations. In addition, the same observations may have different values when assimilated together with different observations. 

\subsection{Ability to assess the accumulated value of observations}\label{sec:ability to assess the accumulated value of observations}

The operational DA process is typically cycled at regular time intervals, e.g., every 1 h for convection-permitting NWP \citep{MilanEtAl2020}. The forecast produced from the analysis of the current DA cycle is used to create the background for the next DA cycle. This means that the information from the observations assimilated in one DA cycle is carried forward to the following cycles. However, the DFS, PAI, EFSOI, C-V and ESA methods can only be used to assess the value of observations for the analysis and forecast of the same cycle. In other words, they cannot be used to assess the value of the observations assimilated in previous cycles for the analyses and forecasts of the current cycle. In contrast, an OSE, a SOSE, or the EDA method can be carried out in a cycling environment, allowing the assessment of accumulated values of observations. The SOSE implementation is described as a single-cycle experiment, but it has the potential to be applied in a cycling environment \citep{Marseille2008}.

\subsection{Ability to assess the value of anchoring}

In NWP systems, relatively precise observations that can be accurately simulated by the system (e.g., global navigation satellite system radio occultation and radiosonde observations) are used as a reference to minimize the effect of biases in the model and other observations \citep[e.g.,][]{Devon2023,Chandramouli2022,Auligne2007}. Bias correction is important for convection-permitting NWP, as observational and model biases are potentially larger at small scales. To assess the additional value of observations as ``anchor" observations, the method should involve the innovation vector. For example, when using the EFSOI, we can calculate the sensitivity of the bias coefficients to the anchoring sensors, as well as the impact of the bias correction on the EFSOI. Since the updated bias is cycled to the next DA cycle, we may need to calculate the impact of the bias correction in a cycling environment.

\section{Operational considerations and future recommendations}\label{sec:operational considerations}

In this section, we discuss operational considerations for assessing the value of observations in convection-permitting NWP. While focusing on convection-permitting NWP, we also include some general considerations applicable to both global and convection-permitting NWP. We also provide some future recommendations for assessing the value of observations.

\subsection{Ensemble resolution, size and spread}\label{sec:ensemble}

The EFSOI, the EFSR, the C-V diagnostics and the ESA method are suitable for use in convection-permitting NWP for several reasons: 1) they are computationally inexpensive, 2) they are computationally robust to the nonlinearity of error growth, and 3) they are designed specifically for ensemble forecasting systems (assuming that future convection-permitting NWP systems are ensemble based). However, their successful use relies on the availability of suitable ensembles concerning size (number of ensemble members), domain, resolution and output. The resolution of the ensemble should be high enough to resolve the convective-scale atmospheric processes of interest so that we can assess the value of observations for these scales. The ideal ensemble size is not yet clear. In an idealized study, \citet{Griewank2023} found that the ensemble variance reduction calculated using the ESA method reached stability when the ensemble size was about an order of magnitude smaller than the number of state variables. However, the ensemble size of current operational regional ensemble forecasting systems is typically less than 100 \citepalias{TIGGE-LAM}, which is many orders of magnitude smaller than the number of state variables. In practice, we may consider using a statistical approach to correct the sampling error caused by an insufficient ensemble size \citep{Necker2020}, which may improve the accuracy of the sensitivity vector. 

The ensemble spread can greatly affect the calculation of the sensitivity vector in the ESA method; an ensemble spread that is too small can substantially inflate the raw sensitivity \citep{Ancell2022}. This can be seen from Eq. \eqref{eq:ESA sensitivity vector diagonal}. In addition, if the EDA spread is under dispersed \citep{Bonavita2016}, then the EDA spread reduction may underestimate the absolute value of observations on forecast skill. However, the under-dispersion may not be an issue when comparing the relative impact of two observing strategies. A simple solution to the under-dispersion is to apply a scaling factor to the raw EDA spread reduction. This scaling factor may be different from the covariance inflation factor used for operational forecasting. How the scaling factor should vary with the observing system requires further research. Another solution is to improve the reliability of the EDA by better modelling the sources of uncertainty, such as improving model uncertainty parameterizations and better modelling observation errors.

Easily accessible regional ensemble forecast datasets \citepalias[e.g.,][]{TIGGE-LAM} may be useful for observation impact studies. For example, when using the ESA method, we only need to know the background ensemble perturbation matrix if the sensitivity vector, the observation operator and the observation uncertainty are known (see section \ref{sec:ESA}). However, when using regional ensembles, we should bear in mind that different ensembles will be optimized for the weather type of their region. Consequently, observation impact studies conducted in one region may not be informative for other regions. 

\subsection{Covariance localization}\label{sec:covariance localization}

In ensemble-based DA, spatial covariance localization is mainly used to remove spurious long-range error correlations between model variables when a small ensemble (that is much smaller than the model dimension) is used to estimate these correlations \citep{Houtekamer2001,hamill2001}. Covariance localization also brings some computational side benefits such as the increase in effective ensemble size \citep{oke2007} and the decrease in computational costs \citep{petrie2010}. We should include the same localization when assessing the value of observations using the PAI, EFSOI, EFSR, C-V and ESA methods. In addition to the covariance localization, the covariance inflation used in ensemble-based DA can also affect the estimation of the value of observations \citep{Hotta2021,Kotsuki2019}.

\subsubsection{Flow-dependent localization}

The impact of assimilated observations will evolve as the forecast proceeds, and this evolution should be reflected in the localization function \citep{Kalnay2012}. How the localization function should evolve depends on the observations to be assessed and the atmospheric processes at the time of assessment \citep{Griewank2023}. In general, a flow-dependent localization is needed, which should be as consistent as possible with the dynamical flow \citep{Bocquet2016}. The idea of using flow-dependent localization in the assessment of observation impact is the same as the idea of using flow-dependent localization in DA. In a DA system such as 4DEnVar, we need to calculate cross-covariances between perturbations at different times within the assimilation window. A fixed-in-time localization of the cross-covariance matrices is inappropriate, as the localization should adapt to variations in the cross-covariance matrices \citep{Bishop2007,Desroziers2016}. Simply speaking, when assessing the impact of observations, different localizations are needed to assess the impact on forecasts at different lead-times. Strictly speaking, a variable-dependant localization may also be needed as tracer information (e.g., humidity and hydrometeors) will be advected while waves (wind and temperature) propagate differently.

Several studies have addressed flow-dependent localization in the calculation of the EFSOI. \citet{Kalnay2012} has shown that observation impact estimates could be improved by simply displacing the localization function with a mean group velocity. \citet{Ota2013} found that moving the center of the localization function proportionally to the average of the analysis and forecast horizontal wind at each vertical level can account for the effect of the propagation of the observation impact by the mean flow. \citet{Gasperoni2015} employed a Monte Carlo “group filter” technique to estimate the optimal localization from a large ensemble and found remarkable improvement for longer forecast times and at mid-latitudes. The exact calculation of the time evolution of the localization function in the computation of the sensitivity vector is not an easy task. For an operational NWP model, a precise calculation is not feasible due to the complexity and nonlinearity of model dynamics \citep{Griewank2023}. 

The importance of flow-dependent localization may depend on forecast lead-times and atmospheric processes \citep{Sommer2014}. Nevertheless, flow-dependent localization may be important for quantifying the impact of observations in convection-permitting NWP, and further research is needed.

\subsubsection{The effect of localization on the DFS}\label{sec:the effect of localization on the DFS}

\citet{Hotta2021} have shown that in optimal DA systems, an upper bound of the DFS can be given as 
\begin{equation}
    \e{DFS}\leq\e{min}\{\e{rank}(\m{R}),\e{rank}(\m{H}),\e{rank}(\m{B})\}\leq\e{min}\{m,n\},
\end{equation}
where $m$ denotes the number of observations assimilated and $n$ denotes the dimension of the model state vector. Furthermore, if the DFS is computed with any ensemble-based DA schemes that use $N$ ensemble members to estimate the background error covariance matrix, then $\e{rank}(\m{B})\leq N-1$, and hence \citep{Hotta2021}
\begin{equation}\label{eq:DFS upper bound N-1}
    \e{DFS}\leq N-1,
\end{equation}
which indicates that the DFS can be underestimated if the ensemble size is too small. Applying covariance localization to the background error covariance matrix can increase its rank, and therefore the upper bound of the DFS given by Eq. \eqref{eq:DFS upper bound N-1} increases to \citep{Hotta2021} 
\begin{equation}
    \e{DFS}\leq N\cdot\e{rank}(\m{L}_{nn})-1,
\end{equation}
where $\m{L}_{nn} \in \reals^{n \times n}$ is the model-space localization matrix. 

\subsection{Simulation of observations}\label{sec:simulation of observations}

Synthetic observations are used when assessing the value of observations that are not currently available. These simulated observations should contain realistic observation errors and also have realistic spatial distribution and temporal frequency \citep{Prive2021_Tellus}. The observation types required for convection-permitting NWP may require the consideration of complex errors, such as inter-channel correlated errors and spatially and temporally correlated errors \citep{Janjic2018}, which complicates simulating these observations meaningfully \citep{Hoffman2016}. 

Simulating the spatial and temporal distribution of observations is also a challenge. In an OSSE (see section \ref{sec:OSSE}), an NR of an NWP model is defined as the truth. When simulating observations, the NR is first interpolated to the times and locations of the real observations, and then the simulated observation errors are added. To capture convection-permitting scales, the NR needs to be available at both a very high spatial resolution and a very high temporal frequency. For NR with very high spatiotemporal resolution, the input/output and storage requirements can be very large, especially for global models or regional models covering a large area. Alternative approaches to generating the NR and synthetic observations may be necessary depending on available computational and storage resources. There is an additional layer of complexity in that to make full use of very high-resolution NR: some observations should be treated differently than they have been in the past. For example, the footprint of radiance observations should be considered during the simulation, rather than doing a simple point interpolation of the NR.

Since it is not possible to produce an exact simulacrum of real observations, decisions must be made about what aspects of the real observations need to be faithfully represented by the synthetic observations. This may require testing and experimentation to determine what characteristics of observations are important when working with convection-permitting DA.

\subsection{Verification reference}\label{sec:verification reference}

The uncertainty in the verification reference could increase the uncertainty in the assessment of the value of observations, and ideally, errors in the verification reference should be unbiased and statistically independent from the forecast errors \citep{Daescu2009}.

Because the analysis provides complete and uniform spatial coverage, it is a commonly used forecast validation reference for assessing the impact of observations on global forecasts. The analysis is also, by definition, the optimal estimate of the atmospheric state. However, in convection-permitting NWP, the analysis may contain non-negligible systematic errors \citep{Necker2018}. More importantly, forecast and analysis errors are unlikely to be independent for short-range forecasts \citep[e.g., a few hours;][]{Prive2021}. The positive correlation between the forecast error and the analysis error means that forecast skill is overestimated \citep{Hotta2023}. Using the analysis from a different NWP system may remove the error correlations, but it may introduce other problems, such as inconsistency between model grids and reduced accuracy in the analysis. We may only want to verify the forecasts using analysis from an NWP system of better or similar quality. Using the analysis produced by an independent run with the same NWP system can reduce the error correlation but adds computational cost. Practical approaches to generating twin-analysis in an operational environment are required \citep{Hotta2023}.

Compared to the analysis, observations are independent of model forecasts and can be a better choice for verifying the forecast in convection-permitting NWP. In addition, the observations can also be used to evaluate the accuracy of the analysis. However, there are still challenges in using observations as the verification reference. First, observations can contain much larger random and systematic errors than the analysis. Second, observations can be sparse and inhomogeneous in space and time. Third, we need appropriate observation operators that calculate the anticipated values of some types of observations (e.g., satellite radiances) for a given model state. The error in the observation operator affects the verification. The key point in using observations as the verification reference is to choose observations that are well distributed in space and time and that represent the domain of interest and the (most) critical atmospheric variables for assessing a successful forecast. For convection-permitting forecasts, the atmospheric variables that are of most interest may include precipitation, wind gusts, surface wind and temperature, and total hours of sunshine \citep{Necker2018}. 

In an OSSE, the truth is fully known, so model fields can be compared directly to the NR to determine analysis and forecast errors, and observation impacts can be calculated without the issue of self-analysis verification. However, to verify convection-permitting forecasts, we need an NR that can realistically reproduce convective scales, especially when the NWP model has drifted into its own preferred climatology. This NR needs to be carefully validated against the real world, especially for behaviour at small scales in the geographical area of interest. Such validation requires datasets that have spatiotemporal distributions sufficient to observe this type of behaviour, similar to the challenges of verifying forecasts using observations. In addition, some of the issues may be similar to the current challenges in verifying precipitation, such as distinguishing displacement, magnitude, and form of features. How the validation of the NR would occur, and what datasets would be available to affect such a validation, is the major issue. 

\subsection{Verification metric}\label{sec:verification metric}

Different choices of verification metrics and forecast lead-times may strongly affect estimates of the impact of observations. In convection-permitting NWP, we need to carefully select verification metrics that are suitable for small-scale features and can provide meaningful information. For instance, we are more interested in forecast skill for high-impact weather events, such as heavy precipitation and tornadoes. The metrics used for these events should provide spatial information \citep{DeyPhDThesis,Ebert2008} and measure multiple aspects such as intensity, location, timing and structure \citep{Gilleland2009}. For the evaluation of ensemble forecasts, we also need to consider the effect of the ensemble size on the verification metric \citep{Clark2011} and combine ensemble information with spatial information \citep{Dey2016SpatialPrecipitation,Dey2016New}. In addition to measuring the forecast skill of a weather event, the overall forecast skill can be summarised by using a weighted average of skill scores for a group of variables at different forecast times \citep[such as the Met Office UK NWP index;][]{Simonin2019}. 

In an OSSE, the choice of verification metrics ties into the validation of the NR (see section \ref{sec:verification reference}), and also the validation of the performance of the OSSE framework as a whole. Ideally, the convective-scale model error growth in an OSSE would be representative of the error growth in the real world. However, \citet{Yu2019} have shown that even though an OSSE may have realistic error growth rates, this is not necessarily sufficient to achieve accurate observation impacts.

\subsection{Lateral boundary conditions}\label{sec:lateral boundary conditions}

Lateral boundary conditions may have a large effect on forecast skill \citep[][]{Gustafsson1998}, thus the estimated impact of observations can be different when different lateral boundary conditions are given. The effect of the lateral boundary condition on the estimation of the observation impact may need to be explicitly considered. Furthermore, the observations used in global NWP affect the accuracy of the lateral boundary condition provided to a LAM, and thus also impact the forecast skill of the LAM. In some cases, improvements in boundary conditions may have a larger impact on forecast skill than the regional assimilation of some observations does \citep{Atlas2015}. In this case, the value of globally assimilated observations for regional forecasts may need to be taken into account \citep{Milan2023}.

Considerations should also be given to the provision of appropriate lateral boundary conditions for a regional OSSE. A proper regional OSSE will have a regional NR embedded within a global NR, which is a challenging enterprise that is rarely done \citep{Atlas2015}. If a regional NR is used with boundary conditions from a “real world” forecast of a global model \citep{Duruisseau2017}, then one is constrained to looking at limited case studies, with the forecast length limited in part by the time it takes for information to propagate from the boundary conditions. This type of NR is best suited to localized observing networks and analysis or very short forecasts, which is mostly what we are interested in for convection-permitting NWP.

\subsection{Retuning of the background error covariance matrix}\label{sec:retuning of the background error covariance}

When carrying out an OSE, an OSSE or a SOSE, an open question is whether we should retune the background error covariance matrices for each combination of assimilated observations \citep{WMOreport,Duncan2021}. Theoretically, changes in observations will affect the estimation of the background error covariance matrix. For example, when using the EDA to estimate the background error covariance matrix, changes in observations will alter the EDA spread and hence the resultant matrix. Practically, \cite{Duncan2021} found that the effect of retuning the background error covariance matrices is small and should not change the conclusions drawn from a typical OSE (without retuning the background error covariance matrix). Given the additional computational cost of re-running the EDA, it is worth discussing whether the background error covariance matrix needs to be updated, and which kind of changes to the observing system might require the retuning of the background error covariance matrix to ensure reliable conclusions. In low-baseline experiments, where large proportions of observations are to be denied, the use of the background error covariance matrix obtained from the full observing system may be more problematic \citep{kelly2008impact}. 

\section{Summary}\label{sec:summary}

\begin{table}
    \footnotesize
    \centering
     \setlength{\leftmargini}{0.4cm}
    \begin{tabular}{ | m{1.2cm} | m{6.6cm} | m{6.6cm} |}
        \hline
        Methods & Key remarks & Future research directions\\
        \hline
        DFS & 
        \begin{itemize} 
            \item A widely used observation influence method.
        \end{itemize} & 
        \begin{itemize} 
            \item Calculating the DFS in ensemble forecasting systems.
            \item Covariance localization and inflation.
        \end{itemize} \\
        \hline
        PAI & 
        \begin{itemize} 
            \item A new 3-D observation influence method for ensemble-based DA systems.
        \end{itemize} & 
        \begin{itemize} 
            \item Application and evaluation.
            \item Comparison to other methods.
            \item Use as a diagnostic tool to optimize ensemble-based DA systems (e.g., localization).
        \end{itemize} \\ 
        \hline
        OSE & 
        \begin{itemize} 
            \item A widely used method for assessing the impact of existing observations.
        \end{itemize} & 
        \begin{itemize} 
            \item Forecast verification reference (analyses or observations).
            \item Retuning of the background error covariance matrix.
        \end{itemize} \\
        \hline
        OSSE & 
        \begin{itemize} 
            \item A widely used method for assessing the impact of future observations.
            \item Performance depends on the quality of the OSSE system.
        \end{itemize} & 
        \begin{itemize} 
            \item Generating and validating the NR.
            \item Simulating high-resolution observations with complicated error statistics. 
            \item Lateral boundary conditions for a regional OSSE.
        \end{itemize} \\
        \hline
        SOSE & 
        \begin{itemize} 
            \item A less common method for assessing the impact of future observations.
        \end{itemize} & 
        \begin{itemize} 
            \item Application and evaluation.
            \item Comparison to other methods. 
        \end{itemize} \\
        \hline
        EFSOI & 
        \begin{itemize} 
            \item A computationally inexpensive alternative to an OSE for ensemble forecasting systems.
        \end{itemize} & 
        \begin{itemize} 
            \item Forecast verification reference (analyses or observations).
            \item Flow-dependent localization.
            \item Effect of the size, spread and resolution of the ensemble.
        \end{itemize} \\
        \hline
        EFSR & 
        \begin{itemize} 
            \item A relatively new method based on the EFSOI for assessing the impact of observation error covariances.
        \end{itemize} & 
        \begin{itemize} 
            \item Application and evaluation.
        \end{itemize} \\
        \hline
        C-V  & 
        \begin{itemize} 
            \item New diagnostics based on the EFSOI for identifying the suboptimality in the assimilation of observations.
        \end{itemize} & 
        \begin{itemize} 
            \item Application and evaluation.
        \end{itemize} \\
        \hline
        EDA & 
        \begin{itemize} 
            \item A method used mainly in global NWP to assess the impact of future observations.
            \item Relying on the spread-skill relationship.
        \end{itemize} & 
        \begin{itemize} 
            \item Application and evaluation in convection-permitting NWP.
            \item Dealing with the under dispersion of the EDA spread.
        \end{itemize} \\
        \hline
        ESA & 
        \begin{itemize}
            \item A computationally efficient method for assessing the impact of observations in ensemble forecasting systems.            
            \item Relying on the spread-skill relationship.
        \end{itemize} & 
        \begin{itemize} 
            \item Effect of the size, spread and resolution of the ensemble.
            \item Flow-dependent localization. 
        \end{itemize} \\
        \hline
    \end{tabular}
    \caption{Key remarks and future research directions for methods to assess the value of observations in convection-permitting NWP.} \label{tab:summary}
\end{table}

Assessing the value of observations in convection-permitting NWP can be challenging due to factors such as strong nonlinearities in the forecast model and observation operator, the LAM domain, and the complexity of the observation types required. These factors narrow the list of methods we can use, increase the difficulty of obtaining statistically significant estimates of the value of observations, and complicate the simulation of observations. To provide guidance for the assessment, we have compared existing methods in terms of their ability to address the described challenges, discussed issues that should be considered in their operational use and identified future research directions. Some general research directions are the choice of using observations or analyses for forecast verification, the simulation of high-resolution observations with complicated error statistics, flow-dependent localization for ensemble-based observation impact methods, and the effect of ensemble size on assessing the value of observations. Specific research directions for each method are summarized in Table \ref{tab:summary}, together with key comments for each method.

We need to be mindful of several things when assessing the value of observations in convection-permitting NWP. The first thing to note is that different methods may have a different focus for use. For example, an OSSE, the EDA method and the ESA method are specifically used to assess the impact of future observations. The DFS, PAIs, C-V diagnostics and EFSR are particularly suitable for identifying the problems with the assimilation system and assessing changes to the assimilation system. An OSE assesses actual improvement (or degradation) in forecast skill and analysis quality in operational NWP. Therefore, its result is often used to validate the results from other methods. Decisions on which methods to use should be based on the user's specific purpose and the resources available at the NWP centre. For example, establishing and maintaining an OSSE system is no easy task. In general, it is sensible to use more than one method to avoid the misinterpretation of the result from any single method. This is particularly important when using synthetic observations to assess the value of future observations.

The second thing to note is that the same observations can have different values in different convection-permitting NWP systems. Convection-permitting NWP typically uses a LAM, which can vary considerably between tropical and mid-latitude regions due to differences in the dominant atmospheric processes \citep{Prive2021, Gelaro2010}. Methods for assessing the value of observations are usually operationally oriented and provide us with statistical estimates of the value of observations. However, we may need to understand the underlying dynamics that cause the results of these methods. Furthermore, the quality of convection-permitting NWP systems can vary considerably due to factors such as differences in local observation networks and differences in NWP models. A small added value should be expected when adding observations to a convection-permitting NWP system that already has good quality and assimilates extensive observations.

To summarise, appropriate selection and sensible use of the available methods, and careful interpretation of the results, can give us a reliable assessment of the value of observations, which can help us to improve the observing system and the DA system, and thus improve the forecast skill in convection-permitting NWP. 

\section*{Acknowledgement}

We acknowledge the discussions that led to the paper via WMO (World Meteorological Organization) WWRP (World Weather Research Programme) DAOS (Data Assimilation and Observing Systems), PDEF (Predictability, Dynamics and Ensemble Forecasting) and NMR (Nowcasting and Mesoscale Research) working groups. 

G. Hu was funded in part by the NERC (Natural Environment Research Council) NCEO (National Centre for Earth Observation) and Met Office. S. L. Dance was funded in part by the NERC NCEO. U. L\"{o}hnert has been supported by the Hans-Ertel-Centre for Weather Research funded by the German Federal Ministry for Transportation and Digital Infrastructure (Grant number BMVI/DWD 4818DWD5B). X. Wang is supported by NA19OAR0220154.

\section*{Conflict of interest}
We declare we have no competing interests.

\section*{Data availability statement}

No data was used or created.

\bibliographystyle{abbrvnat}
\bibliography{ref}

\begin{appendix}

\renewcommand{\theequation}{A.\arabic{equation}}
\setcounter{equation}{0}
\section{Different formulations for the DFS}\label{appendix:DFS A}

In this appendix, we present several different expressions for the DFS. They are all equivalent to Eq. \eqref{eq:DFS HK}. In variational DA systems, the DFS can be estimated by \citep{RODGERS1998,Fisher2003}
\begin{equation}\label{eq:DFS definition eigenvalues}
    \e{DFS}=n-\sum_{i=1}^n\lambda_i\left(\m{U}\m{A}\m{U}^\top\right),
\end{equation}
where $\m{U}$ is a transform matrix that satisfies $\m{U}\m{U}^\top=\m{B}^{-1}$ (e.g., $\m{U}=\m{B}^{-1/2}$), $\m{U}\m{A}\m{U}^\top$ the analysis error covariance matrix of transformed variable $\m{U}\m{x}_a$ \citep{Courtier1998} and $\lambda_i(\cdot)$ the $i$-th eigenvalue of a matrix. In Eq. \eqref{eq:DFS definition eigenvalues}, $n$ represents the total number of statistically independent directions, and the sum of eigenvalues estimates the number of directions that are not constrained by the observations \citep{Fisher2003}. Since the trace of a matrix is equal to the sum of its eigenvalues, Eq. \eqref{eq:DFS definition eigenvalues} becomes
\begin{equation}\label{eq:DFS definition Fisher}
    \e{DFS}=n-\e{tr}\left(\m{U}\m{A}\m{U}^\top\right).
\end{equation}
Substituting \citep{Bouttier2002}
\begin{equation}
    \m{A}=\m{B}-\m{KHB}
\end{equation}
into Eq. \eqref{eq:DFS definition Fisher} and using the properties of the trace of a sum and product of matrices \citep[][section 2.2]{MatrixMathematics}, we obtain Eq. \eqref{eq:DFS HK}.

The DFS can also be defined as the statistical expectation of the normalized analysis-minus-background squared residuals \citep{Fowler2020}
\begin{equation}\label{eq:DFS definition Fowler}
    \e{DFS}=\expt{\left(\m{x}_a-\m{x}_b\right)^\top\m{B}^{-1}\left(\m{x}_a-\m{x}_b\right)},
\end{equation}
measuring the expected influence of observations in pulling the analysis away from the background. The derivation from Eq. \eqref{eq:DFS definition Fowler} to Eq. \eqref{eq:DFS HK} can be found in \citet[][Appendix A]{Fowler2020}.

Furthermore, the DFS can also be expressed as the trace of the derivative of analysis in observation space with respect to observations \citep{Cardinali2004,Chapnik2006}
\begin{equation}\label{eq:DFS self-sensitivity}
    \e{DFS}=\e{tr}\left[\frac{\partial(\m{H}\m{x}_a)}{\partial \m{y}}\right]=\sum_{i=1}^{m}\frac{\partial(\left[\m{H}\m{x}_a\right]_i)}{\partial y_i}.
\end{equation}
In other terms, the DFS can be considered as the self-sensitivity of analysis with respect to observations (i.e., the influence of the $i$-th observation on the analysis at the $i$-th location). Since we have $\m{x}_a=\m{x}_b+\delta\m{x}$, where $\delta\m{x}$ is given by Eq. \eqref{eq:analysis increment}, we can easily prove the equivalence between Eq. \eqref{eq:DFS self-sensitivity} and Eq. \eqref{eq:DFS HK}.

\renewcommand{\theequation}{B.\arabic{equation}}
\setcounter{equation}{0}
\section{Practical calculations of the DFS}\label{appendix:DFS B}

As described in section \ref{sec:DFS}, Eq. \eqref{eq:DFS HK} cannot be used directly to calculate the DFS in practice. A possible approach is to calculate the matrix-vector product, $\m{HK}$, using a low-rank approximation of the background error covariance matrix \citep{Cardinali2004,Fisher2003}. In addition, we describe two approaches in this appendix: the first uses a sample of assimilation residuals \citep{Fowler2020}, and the second is derived based on a randomized estimation of matrix trace and exploits observation and analysis perturbations \citep{Desroziers2001,Desroziers2005,Wahba1995}.

\renewcommand{\theequation}{B.1.\arabic{equation}}
\setcounter{equation}{0}
\subsection{The assimilation residual approach}

\citet{Fowler2020} has shown that Eq. \eqref{eq:DFS HK} can be estimated by
\begin{equation}\label{eq:DFS residuals}
    \e{DFS}=\e{tr}\left(\expt{\m{R}^{-1/2}\m{d}^a_b\left(\m{R}^{-1/2}\m{d}^o_a\right)^\top}\left(\expt{\m{R}^{-1/2}\m{d}^o_b\left(\m{R}^{-1/2}\m{d}^o_a\right)^\top}\right)^{-1}\right),
\end{equation}
where 
\begin{equation}
    \m{d}^a_b=H(\m{x}_a)-H(\m{x}_b)
\end{equation}
is a vector containing the analysis increments in the observation space. The expected value is estimated by averaging over a sample of $\m{d}^o_a$, $\m{d}^o_b$ and $\m{d}^a_b$ obtained from cycling the DA system. Similarly, Eq. \eqref{eq:DFS actual} can be calculated as \citep[][]{Fowler2020,Lupu2011}
\begin{equation}\label{eq:DFS actual residuals}
    \e{DFS}_{Actual}=\expt{\left(\m{d}_a^o\right)^\top\m{R}^{-1}\m{d}^a_b}.
\end{equation}
We note that Eqs. \eqref{eq:DFS residuals} and \eqref{eq:DFS actual residuals} were derived by assuming a linear observation operator. To take into account the nonlinearity of the observation operator, modifications should be made to these two equations (see the supporting information for \citet{Fowler2020}). The assimilation residual approach has been used in variational DA systems \citep{Fowler2020,Lupu2011}, but they should also be applicable to ensemble-based DA (e.g., LETKF).

\renewcommand{\theequation}{B.2.\arabic{equation}}
\setcounter{equation}{0}
\subsection{The randomized approach}

Let $\m{G}$ be a square matrix, then an unbiased estimate of $\e{tr}(\m{G})$ is \citep{girard1989}
\begin{equation}\label{eq:randomized trace estimate}
    T_{\m{G}}(\boldsymbol{\eta})=\boldsymbol{\eta}^\top\m{G}\boldsymbol{\eta},
\end{equation}
where $\boldsymbol{\eta}$ is a random vector whose elements satisfy $\expt{\eta_i}=0$, $\expt{\eta_i^2}=1$ and $\expt{\eta_i\eta_j}=0$ for $i\neq j$. The mean value of $T_{\m{G}}(\boldsymbol{\eta})$ is  
\begin{equation}
    \expt{T_{\m{G}}(\boldsymbol{\eta})}=\e{tr}(\m{G}).
\end{equation}
If the elements of $\boldsymbol{\eta}$ are independent random values from the standard Gaussian distribution, i.e., $\boldsymbol{\eta} \sim \mathcal{N}(\m{0},\m{I})$, then the variance of $T_{\m{G}}(\boldsymbol{\eta})$ is \citep{girard1989}
\begin{equation}
    2\sum_i\sum_jG_{i,j}^2.
\end{equation}
Now, we rewrite Eq. \eqref{eq:DFS HK} as 
\begin{equation}
    \e{DFS}=\e{tr}(\m{R}^{1/2}\m{R}^{-1}\m{HK}\m{R}^{1/2}).
\end{equation}
Then, using Eq. \eqref{eq:randomized trace estimate} we obtain \citep{Lupu2011,Desroziers2005,Desroziers2001,Wahba1995},
\begin{equation}\label{eq:DFS Desroziers2001}
    \e{DFS}=\expt{\delta \m{y}^\top\m{R}^{-1}\m{H}\delta\m{x}(\delta\m{y})},
\end{equation}
where 
\begin{equation}
    \delta \m{y}=\m{R}^{1/2}\boldsymbol{\eta}    
\end{equation}
is a vector of observation perturbations and 
\begin{equation}
    \delta\m{x}(\delta\m{y})=\m{K}\delta\m{y}
\end{equation}
is the difference between the analysis increments obtained with perturbed and unperturbed observations. The randomized approach is readily applicable to stochastic ensemble Kalman filters \citep[e.g.,][]{Houtekamer2001} where the observation vector, $\m{y}$, is perturbed and each ensemble member is updated with an identical Kalman gain matrix, $\m{K}$. The randomized approach is also applicable to EDA \citep{Desroziers2009}. Assuming that $\partial\left[H(\m{x}_a)\right]/\partial\m{y}$ varies smoothly with respect to observations, Eq. \eqref{eq:DFS Desroziers2001} can be used when the observation operator is nonlinear \citep{Chapnik2006}. There is also a randomized approach which perturbs the analysis control vector rather than the observation vector \citep{Fisher2003}.

\renewcommand{\theequation}{C.\arabic{equation}}
\setcounter{equation}{0}
\section{Derivation of the EFSOI}

The EFSOI can be calculated using self-analysis verification or observation-based verification.

\renewcommand{\theequation}{C.1.\arabic{equation}}
\setcounter{equation}{0}
\subsection{Self-analysis verification}\label{appendix:EFSOI derivation}

Eq. \eqref{eq:delta e} is an expression of the difference between two squares that can be factorised as \citep{Kalnay2012}
\begin{align}\label{eq:difference of two squares factorisation}
    \begin{split}
        \delta e &= \left[\left(\m{x}^f_a-\m{x}_t\right) - \left(\m{x}^f_b-\m{x}_t\right)\right]^\top\m{C}\left[\left(\m{x}^f_a - \m{x}_t\right) + \left(\m{x}^f_b - \m{x}_t\right)\right]\\
        &= \left(\delta\m{x}^f\right)^\top\m{C}\left[\left(\m{x}^f_a - \m{x}_t\right) + \left(\m{x}^f_b - \m{x}_t\right)\right],
    \end{split}
\end{align}
where $\delta\m{x}^f=\m{x}^f_a-\m{x}^f_b$ is difference between two forecasts. The forecast difference can be written as a linearized function of the analysis increments, 
\begin{equation}\label{eq:linearised forecast model}
    \delta \m{x}^f\approx\m{M}\delta\m{x},
\end{equation}
where $\m{M}$ denotes a tangent linear model of the forecast model linearised about the initial state or a perturbation forecast model. Sequentially substituting Eq. \eqref{eq:linearised forecast model} and Eq. \eqref{eq:analysis increment} into Eq. \eqref{eq:difference of two squares factorisation} gives an expression for the FSOI,
\begin{equation}\label{eq:FSOI}
    \delta e\approx\left(\m{d}^o_b\right)^\top\m{K}^\top\m{M}^\top\m{C}\left[\left(\m{x}^f_a-\m{x}_t\right)+\left(\m{x}^f_b-\m{x}_t\right)\right].
\end{equation}
(There is another expression for the FSOI that uses two different tangent linear models for background and analysis trajectories respectively \citep{Errico2007,Langland2004}). Substituting \eqref{eq:K ensemble estimate} into Eq. \eqref{eq:FSOI} and assuming
\begin{equation}\label{eq:EFSOI linear assumption}
    \m{X}^f_a \approx \m{M}\m{X}_a,
\end{equation}
we obtain the expression for the EFSOI (Eq. \eqref{eq:EFSOI}). The calculation of Eq. \eqref{eq:EFSOI} relies on the DA system producing ensemble perturbations of the analysis in the observation space. Such ensemble perturbations are not normally produced in the 4DEnVar system, where the EFSOI needs to be calculated differently \citep{Buehner2018}. Applying covariance localization to Eq. \eqref{eq:EFSOI} gives
\begin{equation}\label{eq:EFSOI localization}
    \e{EFSOI} = \frac{1}{N-1}\left(\overline{\m{d}}^o_b\right)^\top\m{R}^{-1}\left(\m{L}_{mn}\circ\m{Y}_a\left(\m{X}_a^f\right)^\top\right)\m{C}\left[\left(\overline{\m{x}}^f_a - \m{x}_t\right) + \left(\overline{\m{x}}^f_b - \m{x}_t\right)\right],
\end{equation}
where $\circ$ denotes the element-wise product and $\m{L}_{mn} \in \reals^{m \times n}$ the corresponding observation-model localization used in the assimilation. It should be noted that including localization in the EFSOI as Eq. \eqref{eq:EFSOI localization} may introduce large errors. This is because the Kalman gain matrix, $\m{K}$, is propagated by the tangent linear model, $\m{M}$, and if the matrix $\m{K}$ is localized, then $\m{MK}$ needs a propagated localization. Mathematically, $\m{M}(\m{L}_{nm}\circ\m{K}) \neq \m{L}_{nm}\circ(\m{MK})$. For more information on the propagation of localization see \citet{Griewank2023}.

\renewcommand{\theequation}{C.2.\arabic{equation}}
\setcounter{equation}{0}
\subsection{Observation-based verification}\label{appendix:observation-based verification}

When observations are used to verify the forecast \citep[e.g., radar-derived precipitation observations;][]{Necker2018}, Eq. \eqref{eq:delta e} becomes
\begin{equation}\label{eq:difference of two squares y}
    \delta e_y=\left(\overline{H_v(\m{x}^f_a)}-\m{y}_v\right)^\top\m{R}^{-1}_v\left(\overline{H_v(\m{x}^f_a)}-\m{y}_v\right)-\left(\overline{H_v(\m{x}^f_b)}-\m{y}_v\right)^\top\m{R}^{-1}_v\left(\overline{H_v(\m{x}^f_b)}-\m{y}_v\right),
\end{equation}
where $\m{y}_v \in \reals^{m_v}$ is the verifying observation vector, $\m{R}_v \in \reals^{m_v \times m_v}$ the verifying observation error covariance matrix and $H_v: n \rightarrow m_v$ the verifying observation operator that maps model state space into verifying observation space. We note that Eq. \eqref{eq:difference of two squares y} is similar to Eq. \eqref{eq:Todling2013} if the forecast-lead time is set to $0$. Following Eq. \eqref{eq:difference of two squares factorisation}
and using a linearised $H_v$, we obtain
\begin{equation}\label{eq:difference of two squares factorisation y}
    \delta e_y \approx \left(\m{H}_v\delta\m{x}^f\right)^\top\m{R}_v^{-1}\left[\left(\overline{H_v(\m{x}^f_a)}-\m{y}_v\right)+\left(\overline{H_v(\m{x}^f_b)}-\m{y}_v\right)\right].
\end{equation}
Substituting Eq. \eqref{eq:linearised forecast model}, Eq. \eqref{eq:analysis increment} and Eq. \eqref{eq:K ensemble estimate} into Eq. \eqref{eq:difference of two squares factorisation y}, we obtain the observation-based EFSOI,
\begin{equation}\label{eq:EFSOI observation}
    \e{EFSOI}_y = \frac{1}{N - 1}\left(\overline{\m{d}}^o_b\right)^\top\m{R}^{-1}\m{Y}_a\left(\m{Y}_{av}^f\right)^\top\m{R}_v^{-1}
\left[\left(\overline{H_v(\m{x}^f_a)} - \m{y}_v\right) + \left(\overline{H_v(\m{x}^f_b)} - \m{y}_v\right)\right],
\end{equation}
where $\m{Y}_{av}^f \in \reals^{m_v \times N}$ is the analysis forecast ensemble perturbation matrix in verifying observation space and we assume $\m{Y}_{av}^f=\m{H}_v\m{X}_a^f$. Covariance localization can be applied to the observation-based EFSOI in a similar way as Eq. \eqref{eq:EFSOI localization}. 

In addition to Eq. \eqref{eq:EFSOI observation}, \citet{Sommer2016} suggested calculating the observation-based EFSOI as a linearization around the analysis. As a result, the forecast from the background is no longer needed in the calculation. However, it should be noted that the mathematical expressions from \citet{Sommer2016}'s method are not exactly the same as the ones from the $\e{EFSOI}_y$ given by Eq. \eqref{eq:EFSOI observation}.

\section{The ESA method}\label{appendix:ESA derivation}

\renewcommand{\theequation}{D.\arabic{equation}}
\setcounter{equation}{0}

Let $\m{J} \in \reals^{N}$ be a vector containing a set of $J$ (defined in section \ref{sec:ESA}) computed for each ensemble member, then the variance of $\m{J}$ about the ensemble mean value is
\begin{equation}\label{eq:ESA covariance of J}
    \sigma^2_J=\frac{1}{N-1}\delta \m{J}\delta \m{J}^\top,
\end{equation}
where $\delta \m{J} \in \reals^{N}$ is the ensemble perturbation calculated by subtracting the ensemble mean from $\m{J}$. We may relate $\delta \m{J}$ to ensemble perturbations of model state variables using the first-order Taylor expansion about the ensemble mean,
\begin{equation}\label{eq:ESA sensitivity}
    \delta \m{J} \approx \left[\frac{\partial J}{\partial \m{x}}\right]^\top\m{X}_b.
\end{equation}
Substituting Eq. \eqref{eq:ESA sensitivity} into Eq. \eqref{eq:ESA covariance of J} gives
\begin{equation}\label{eq:delta J pb}
    \sigma^2_J=\left[\frac{\partial J}{\partial \m{x}}\right]^\top\m{P}_b\left[\frac{\partial J}{\partial \m{x}}\right].
\end{equation}
Similarly, we may calculate $\delta \m{J}$ using the analysis perturbation matrix, and obtain
\begin{equation}\label{eq:delta J pa}
    \sigma^2_J=\left[\frac{\partial J}{\partial \m{x}}\right]^\top\m{P}_a\left[\frac{\partial J}{\partial \m{x}}\right].
\end{equation}
Subtracting Eq. \eqref{eq:delta J pb} from Eq. \eqref{eq:delta J pa} gives Eq. \eqref{eq:EVR of J}. The ESA method can use either an explicit sensitivity vector or an implicit sensitivity vector \citep{Griewank2023}. 

\renewcommand{\theequation}{D.1.\arabic{equation}}
\setcounter{equation}{0}
\subsection{Implicit sensitivity vector}\label{appendix:ESA derivation implicit}

Substituting \citep{Nichols2010}
\begin{equation} \label{eq:A=B-KHB ensemble}
    \m{P}_a - \m{P}_b = -\m{P}_b\m{H}^\top\left(\m{H}\m{P}_b\m{H}^\top+\m{R}\right)^{-1}  \m{H}\m{P}_b
\end{equation}
into Eq. \eqref{eq:EVR of J} gives
\begin{equation}\label{eq:ESA middle product}
    \delta\sigma^2_J=-\left[\frac{\partial J}{\partial \m{x}}\right]^\top\m{P}_b\m{H}^\top\left(\m{H}\m{P}_b\m{H}^\top+\m{R}\right)^{-1}\m{H}\m{P}_b\left[\frac{\partial J}{\partial \m{x}}\right].
\end{equation}
Right multiplying both sides of Eq. \eqref{eq:ESA sensitivity} by $(N-1)^{-1}\m{X}_b^\top$, we obtain
\begin{equation}\label{eq:implicit sensitivity calculation}
    \left[\frac{\partial J}{\partial \m{x}}\right]^\top\m{P}_b \approx \frac{1}{N-1}\delta \m{J}_b  \m{X}_b^\top.
\end{equation}
Substituting Eq. \eqref{eq:implicit sensitivity calculation} into Eq. \eqref{eq:ESA middle product} gives \citep{Griewank2023,Tardif2022,Hakim2020,Torn2014}
\begin{equation}\label{eq:EVR formula}
    \delta\sigma_J^2=-\frac{1}{(N-1)^2}\left(\delta \m{J}_b\left(\m{Y}_b\right)^\top\right)\left(\frac{1}{N-1}\m{Y}_b\left(\m{Y}_b\right)^\top+\m{R}\right)^{-1}\left(\delta \m{J}_b\left(\m{Y}_b\right)^\top\right)^\top,
\end{equation}
where $\m{Y}_b \in \reals^{m\times N}$ is a matrix containing the background ensemble perturbations in observation space whose $i$-th column is the difference of the $i$-th ensemble member of $H(\m{x}_b)$ from the ensemble mean of $H(\m{x}_b)$ and we assume $\m{Y}_b=\m{H}\m{X}_b$. The calculation of Eq. \eqref{eq:EVR formula} does not require the sensitivity vector. However, if covariance localization is used, Eq. \eqref{eq:EVR formula} becomes
\begin{equation}\label{eq:ESA formula localization}
    \delta\sigma_J^2 = -\frac{1}{(N-1)^2}\left[\frac{\partial J}{\partial \m{x}}\right]^\top\m{L}_{nm}\circ\m{X}_b(\m{Y}_b)^\top\left(\frac{\m{L}_{mm}\circ\m{Y}_b(\m{Y}_b)^\top}{N-1} + \m{R}\right)^{-1}\left(\delta \m{J}_b\m{Y}_b^\top\right)^\top,
\end{equation}
where $\m{L}_{nm} \in \reals^{n \times m}$ is the model to observation localization matrix and $\m{L}_{mm} \in \reals^{m \times m}$ the observation space localization matrix. The calculation of Eq. \eqref{eq:ESA formula localization} requires the sensitivity vector. \citet{Griewank2023} has shown that if the propagated sensitivity localization matrix is known, then we may still omit the sensitivity vector when applying covariance localization in the implicit calculation.

\subsection{Explicit sensitivity vector}\label{appendix:ESA derivation explicit}

\renewcommand{\theequation}{D.2.\arabic{equation}}
\setcounter{equation}{0}

The sensitivity vector can be obtained by solving the linear system given by Eq. \eqref{eq:implicit sensitivity calculation}. However, this linear system is usually ill-conditioned as the matrix $\m{P}_b$ typically has a large condition number. \cite{Nomokonova2022} and \cite{Griewank2023} have shown that Tikhonov regularization \citep{Tikhonov1965} can be used to improve the conditioning of the linear system and enables a direct numerical solution of the sensitivity vector. 

A particular case of solving the linear system is to ignore the non-diagonal elements of $\m{P}_b$, which gives
\begin{equation}\label{eq:ESA sensitivity vector diagonal}
    \left[\frac{\partial J}{\partial \m{x}}\right]_i=\frac{1}{N-1}\cdot\frac{\left[\m{X}_b(\delta\m{J}_b)^\top\right]_i}{\left[X_b\right]_i^2},
\end{equation}
where $i$ is the vector index. However, using a diagonal $\m{P}_b$ to calculate the sensitivity vector will lead to an inaccurate quantitative observation impact estimate \citep{Ren2019}. The sensitivity vector can be thought of as the slope of a linear regression between the scalar metric, $J$, and the state variables \citep{Ancell2022}. If the matrix $\m{P}_b$ is diagonal, then we have a univariate regression. If the matrix $\m{P}_b$ is non-diagonal, then we have a multivariate regression \citep{Hacker2015}.

With an explicit sensitivity vector, $\delta\sigma_J^2$ can be directly computed using Eq. \eqref{eq:EVR of J}, where the analysis ensemble perturbation matrix is calculated by \citep{Griewank2023,Nomokonova2022,Whitaker2002}
\begin{equation}\label{eq:update of analysis perturbations}
    \m{X}_a=\m{X}_b-\whm{K}\m{Y}_b 
\end{equation}
with
\begin{equation}\label{eq:modified Kalman gain matrix}
    \whm{K}=\m{X}_b\left(\m{Y}_b\right)^\top\left(\m{Y}_b\left(\m{Y}_b\right)^\top+\m{R}\right)^{-\top/2}\left(\left(\m{Y}_b\left(\m{Y}_b\right)^\top+\m{R}\right)^{1/2}+\m{R}^{1/2}\right)^{-1}
\end{equation}
being the modified Kalman gain matrix. We can apply the covariance localization in the calculation of $ \m{X}_a$ by replacing $\whm{K}$ with
\begin{equation}\label{eq:localised modified Kalman gain matrix}
    \whm{K}_{loc}=\m{L}_{nm}\circ\m{X}_b\left(\m{Y}_b\right)^\top\left(\m{L}_{mm}\circ\m{Y}_b\left(\m{Y}_b\right)^\top+\m{R}\right)^{-\top/2}\left(\left(\m{L}_{mm}\circ\m{Y}_b\left(\m{Y}_b\right)^\top+\m{R}\right)^{1/2}+\m{R}^{1/2}\right)^{-1}.
\end{equation}
Thus, by using an explicit sensitivity vector, we can easily include the covariance localization into the calculation of $\delta\sigma_J^2$.

\renewcommand{\theequation}{E.\arabic{equation}}
\setcounter{equation}{0}
\section{The C-V diagnostics}\label{appendix:C-V}

The observation-based EFSOI (Eq. \eqref{eq:EFSOI observation}) can be partitioned into two components \citep{Stiller2022},
\begin{equation}\label{eq:efsoi_jajab}
    \e{EFSOI}_y=-(2\beta_{b}-\beta_{ab}),
\end{equation}
where \begin{equation}\label{eq:beta_b}
    \beta_{b} = \left(\overline{\m{d}}^o_b\right)^\top\m{W}\left(\m{y}_v-\overline{H_v(\m{x}_b^f)}\right)
\end{equation}
and
\begin{equation}\label{eq:beta_ab}
    \beta_{ab} = \left(\overline{\m{d}}^o_b\right)^\top\m{W}\left(\overline{H_v(\m{x}_a^f)}-\overline{H_v(\m{x}_b^f)}\right)
\end{equation}
with
\begin{equation}\label{eq:C-V W matrix}
    \m{W} = \frac{1}{N - 1}\m{R}^{-1}\m{Y}_a\left(\m{Y}_{av}^f\right)^\top\m{R}_v^{-1}
\end{equation}
being a cross-covariance matrix. The component $\beta_{b}$ is the key component of the C-V diagnostics, which is the cross-validation of two different observation types (assimilated and verifying observations). 

Assuming we have a linear observation operator, $\m{H}_v$, then
\begin{equation}\label{eq:xaf - xbf Hv}
    \overline{H_v(\m{x}^f_a)} - \overline{H_v(\m{x}^f_b)}\approx \m{R}_v\m{W}^\top\overline{\m{d}}^o_b.
\end{equation}
Using Eq. \eqref{eq:xaf - xbf Hv}, we may rewrite Eq. \eqref{eq:beta_b} as
\begin{equation} \label{eq:beta_b rewrite}
    \beta_{b} = \left(\overline{H_v(\m{x}_a^f)}-\overline{H_v(\m{x}_b^f)}\right)^\top\m{R}_v^{-1}\left(\m{y}_v-\overline{H_v(\m{x}_b^f)}\right)
\end{equation}
and Eq. \eqref{eq:beta_ab} as
\begin{equation} \label{eq:beta_ab rewrite}
    \beta_{ab} = \left(\overline{H_v(\m{x}_a^f)}-\overline{H_v(\m{x}_b^f)}\right)^\top\m{R}_v^{-1}\left(\overline{H_v(\m{x}_a^f)}-\overline{H_v(\m{x}_b^f)}\right).
\end{equation}
Eq. \eqref{eq:beta_b rewrite} shows that $\beta_b$ is positive if the assimilation of observations pulls the model state towards the verifying observations. (A geometric interpretation of Eq. \eqref{eq:beta_b rewrite} is as a scalar product between two vectors. This is positive if the angle between the normalized change of the trajectory in the verifying observation space, $\m{R}_v^{-1/2}\left (\overline{H_v(\m{x}_a^f)}-\overline{H_v(\m{x}_b^f)} \right)$ and the normalized residual vector, $\m{R}_v^{-1/2}\left (\m{y}_v-\overline{H_v(\m{x}_b^f)} \right ) $ is small). Eq. \eqref{eq:beta_ab rewrite} shows that $\beta_{ab}$ is always non-negative and its value depends mainly on the size of analysis increments. 

Based on Eq. \eqref{eq:beta_b} and Eq. \eqref{eq:beta_ab}, the terms $\beta_{b}$ and $\beta_{ab}$ for the observation $j$ are given by 
\begin{eqnarray}
\beta_{b}^{(j)} & = & \left(\m\Pi_{j}\overline{\m{d}}_{b}^{o}\right)^{\top}\m{W}\left(\m{y}_{v}-\overline{H_{v}(\m{x}_{b}^{f})}\right)\label{eq:beta_b-1}\\
\beta_{ab}^{(j)} & = & \left(\m\Pi_{j}\overline{\m{d}}_{b}^{o}\right)^{\top}\m{W}\left(\overline{H_{v}(\m{x}_{a}^{f})}-\overline{H_{v}(\m{x}_{b}^{f})}\right)\label{eq:beta_ab-1}
\end{eqnarray}
where $\m\Pi_{j}$ is the projection operator that sets all vector components to zero, leaving only the component corresponding to the observation $j$ unchanged.

It is interesting to note the similarity between Eq. \eqref{eq:beta_ab rewrite} and the expression for the observation influence in Eq. \eqref{eq:DFS definition Fowler}, which is even more apparent when using the linear approximation to rewrite Eq. \eqref{eq:beta_ab rewrite} as
\begin{equation} 
    \beta_{ab}=\left(\overline{\m{x}_{a}^{f}}-\overline{\m{x}_{b}^{f}}\right)^{\top}\m H_{v}^\top\m{R}_{v}^{-1}\m H_{v}\left(\overline{\m{x}_{a}^{f}}-\overline{\m{x}_{b}^{f}}\right).
\end{equation}
This shows that the term $\expt{\beta_{ab}}$ can be thought of as measuring the expected observation influence in verification rather than model space.

\subsection{Diagnostic 1}\label{appendix:C-V diagnostic 1}
\renewcommand{\theequation}{E.1.\arabic{equation}}
\setcounter{equation}{0}

A negative value of $\e{EFSOI}_y (\approx\delta e_y)$ means that the assimilation of observations reduces the forecast error. Since $\beta_{ab}$ is always non-negative, according to Eq. \eqref{eq:efsoi_jajab}, $\beta_{b} > 0$ is a necessary (but insufficient) condition to ensure a negative $\e{EFSOI}_y$. Furthermore, Eq. \eqref{eq:beta_b rewrite} shows that the sign of $\expt{\beta_{b}}$ can be used to check whether the assimilation of observations pulls the model forecast towards the verifying observations. Therefore, 
\begin{equation}\label{eq:C-V diagnose 1 sign}
    \beta_{b}^{(j)} > 0 
\end{equation}
is a quite fundamental condition for the observation $j$ to have a beneficial contribution to the $\e{EFSOI}_{y}$ by pulling the model state towards the verifying observations. In addition, a more sensitive test is to compare the average (or sum) of $\beta_{b}^{(j)}$ for an observation type with the reference value,
\begin{equation}\label{eq:beta_b tilde}
    \widetilde{\beta}_{b}^{(j)} = \frac{1}{N-1}\e{tr}\left[\m{W}\m{Y}_{bv}^f\left(\m\Pi_{j}\m{Y}_b\right)^\top\right],
\end{equation}
where $\m{Y}_b \in \reals^{m \times N}$ is the background ensemble perturbation matrix in observation space and $\m{Y}_{bv}^f \in \reals^{m_v \times N}$ the background forecast ensemble perturbation matrix in verifying observation space.

The term $\widetilde{\beta}_{b}^{(j)}$ is an ensemble estimate of $\expt{\beta_{b}^{(j)}}$, and it is derived as follows. Using the fact that $\beta_{b}^{(j)} = \e{tr}(\beta_{b}^{(j)})$
and the fact that the trace is invariant under cyclic permutations \citep[][section 2.2]{MatrixMathematics}, we can rewrite Eq. \eqref{eq:beta_b} as 
\begin{equation}\label{eq:C-V Jb trace}
    \beta_{b}^{(j)} = \e{tr}\left[\m{W}\left(\m{y}_v - \overline{H_v(\m{x}_b^f)}\right)\left(\m\Pi_{j}\overline{\m{d}}^o_b\right)^\top\right].
\end{equation}
If all observations and their model equivalents were bias-free and their errors were mutually uncorrelated, then we have 
\begin{equation}\label{eq:estimator}
    \expt{\left(\m{y}_v - \overline{H_v(\m{x}_b^f)}\right)\left(\m\Pi_{j}\overline{\m{d}}^o_b\right)^\top} = \m{H}_v\expt{\err^f_b\left(\err_b\right)^\top}\left(\m\Pi_{j}\m{H}\right)^\top,
\end{equation}
where $\err^f_b \in\reals^n$ is the forecast error vector and $\err_b \in\reals^n$ the background error vector. It should be noted that since the observation error of $\m{y}_v$ is uncorrelated with the observation error of $\m{y}$, there is no observation error cross-covariance term on the right-hand side of Eq. \eqref{eq:estimator}. If the ensemble's capability of representing model error was perfect, then the right-hand side of Eq. \eqref{eq:estimator} can be estimated by 
\begin{equation}\label{eq:ensemble estimator}
    \m{H}_v\expt{\err^f_b\left(\err_b\right)^\top}\left(\m\Pi_{j}\m{H}\right)^\top = \frac{1}{N-1}\m{Y}_{bv}^f\left(\m\Pi_{j}\m{Y}_b\right)^\top,
\end{equation}
which gives $\widetilde{\beta}_{b}^{(j)}$. Thus, the consistency relation, 
\begin{equation}\label{eq:C-V diagnose 1 consistency}
    \expt{\beta_{b}^{(j)}} = \widetilde{\beta}_{b}^{(j)},
\end{equation}
can be used to identify suboptimalities of the assimilation of observations.

\subsection{Diagnostic 2}\label{appendix:C-V diagnostic 2}
\renewcommand{\theequation}{E.2.\arabic{equation}}
\setcounter{equation}{0}

The second term $\beta_{ab}$ is proportional to the size of the analysis increments and can be used to assess to which extent the size of these increments is optimal. Below we show that if the analysis state, $\m{x}_{a}$, is optimal in the sense that $\delta e_{y}$ has a global minimum for this analysis state, then
\begin{equation}\label{eq:C-V diagnose 2}
    \expt{\beta_{ab}^{(j)}} =  \expt{\beta_{b}^{(j)}}
\end{equation}
has to hold. More precisely, we show that if Eq. \eqref{eq:C-V diagnose 2} is not fulfilled we can construct another analysis state, $\wh{\m{x}}_a$, which gives a smaller value of $\delta e_{y}$.

The optimal analysis state is constructed as
\begin{equation} \label{eq:more_optimal_x^a}
    \wh{\m{x}}_a = \m{x}_{a} + \alpha_j\m{K}\m\Pi_{j}\overline{\m{d}}_{b}^{o},
\end{equation}
where 
\begin{equation} \label{eq:delta_lambda}
    \alpha_{j} = \frac{\expt{\beta_{b}^{(j)} - \beta_{ab}^{(j)}}}{\expt{ \beta_{ab}^{(j,j)}}}
\end{equation}
with 
\begin{equation}
    \beta_{ab}^{(j,j)} = \left(\m{H}_v\m{M}\m{K}\m\Pi_{j}\overline{\m{d}}_{b}^{o}\right)^\top\m{R}_v^{-1}\left(\m{H}_v\m{M}\m{K}\m\Pi_{j}\overline{\m{d}}_{b}^{o}\right).
\end{equation}
By definition, $\beta^{(j,j)}_{ab} > 0$. If Eq. \eqref{eq:C-V diagnose 2} is fulfilled, then $\alpha_{j} = 0$ and $\wh{\m{x}}_a = \m{x}_{a}$. Using the linear approximation (which is central to the Kalman filter), the forecast (in verification space) initialized from $\wh{\m{x}}_a$ takes the form,
\begin{equation}
    \overline{H_{v}(\wh{\m{x}}_a^f)} = \overline{H_{v}(\m{x}_{a}^{f})} + \alpha_{j}\m{H}_v\m{M}\m{K}\m\Pi_{j}\overline{\m{d}}_{b}^{o}.
\end{equation}
Substituting this equation into 
\begin{equation}
    \wh{e}_{y,a}=\left(\overline{H_v(\wh{\m{x}}^f_a)} - \m{y}_v\right)^\top\m{R}^{-1}_v\left(\overline{H_v(\wh{\m{x}}^f_a)} - \m{y}_v\right),
\end{equation}
we obtain
\begin{align}
    \begin{split}
        \expt{\wh{e}_{y,a}} 
        &= \expt{e_{y,a}} + \alpha_j^2 \expt{\beta_{ab}^{(j,j)}} - 2\alpha_j \expt{\beta_{b}^{(j)} - \beta_{ab}^{(j)}} \\
        &= \expt{e_{y,a}} - \alpha_j \expt{\beta_{b}^{(j)} - \beta_{ab}^{(j)}} \\
        &= \expt{e_{y,a}} - \frac{\left(\expt{\beta_{b}^{(j)} - \beta_{ab}^{(j)}}\right)^2}{\expt{\beta_{ab}^{(j,j)}}}.
    \end{split}
\end{align}
Thus, we have 
\begin{equation}
    \expt{\wh{e}_{y,a}} < \expt{e_{y,a}}
\end{equation}
for $\alpha_{j} \neq 0$. This expression shows that if Eq. \eqref{eq:C-V diagnose 2} does not hold, then $\wh{\m{x}}_a$ leads to a smaller value of $\delta e_{y}$. Therefore, Eq. \eqref{eq:C-V diagnose 2} (or equivalently $\alpha_{j} = 0$) is a necessary condition for $\m{x}_a$ being a global minimum of $\delta e_{y}$. \citet{Stiller2022} has further shown that Eq. \eqref{eq:C-V diagnose 2} is also a necessary condition for a local minimum at $x = x_a$. A local minimum requires that any derivative of $\delta e_y$ with respect to the initial conditions is zero at $x = x_a$. \citet{Stiller2022} has also shown that if the error covariance matrices employed in the assimilation are completely correct, and there is no bias in the observations and the model states, then the optimal analysis state should be obtained, and Eq. \eqref{eq:C-V diagnose 2} should hold.

\subsection{Diagnostic 3}\label{appendix:C-V diagnostic 3}
\renewcommand{\theequation}{E.3.\arabic{equation}}
\setcounter{equation}{0}

The equivalence relation given by Eq. \eqref{eq:C-V diagnose 1 consistency} can also be used to test the suitability of the ensemble background error covariances by confronting them with error covariances calculated using innovation vectors of two trusted observation types.

\end{appendix}
\end{document}